\documentstyle[12pt]{article}
\input psfig.sty
\def\nn{\noindent}
\def\Re{{\cal R \mskip-4mu \lower.1ex \hbox{\it e}\,}}
\def\Im{{\cal I \mskip-5mu \lower.1ex \hbox{\it m}\,}}
\def\ie{{\it i.e.}}
\def\eg{{\it e.g.}}

\def\etal{{\it et al.}}
\def\ibid{{\it ibid}.}
\def\sub#1{_{\lower.25ex\hbox{$\scriptstyle#1$}}}

\def\to{\rightarrow}

\def\subw{_{\rm w}}
\def\mh{\ifmmode m\sbl H \else $m\sbl H$\fi}
\def\mch{\ifmmode m_{H^\pm} \else $m_{H^\pm}$\fi}
\def\mt{\ifmmode m_t\else $m_t$\fi}
\def\mc{\ifmmode m_c\else $m_c$\fi}
\def\mz{\ifmmode M_Z\else $M_Z$\fi}
\def\mw{\ifmmode M_W\else $M_W$\fi}
\def\mws{\ifmmode M_W^2 \else $M_W^2$\fi}
\def\mhs{\ifmmode m_H^2 \else $m_H^2$\fi}   
\def\mzs{\ifmmode M_Z^2 \else $M_Z^2$\fi}
\def\mts{\ifmmode m_t^2 \else $m_t^2$\fi}
\def\mcs{\ifmmode m_c^2 \else $m_c^2$\fi}
\def\mchs{\ifmmode m_{H^\pm}^2 \else $m_{H^\pm}^2$\fi}
\def\ztwo{\ifmmode Z_2\else $Z_2$\fi}
\def\zone{\ifmmode Z_1\else $Z_1$\fi}
\def\mtwo{\ifmmode M_2\else $M_2$\fi}
\def\mone{\ifmmode M_1\else $M_1$\fi}
\def\tb{\ifmmode \tan\beta \else $\tan\beta$\fi}
\def\xw{\ifmmode x\subw\else $x\subw$\fi}
\def\ch{\ifmmode H^\pm \else $H^\pm$\fi}
\def\lum{\ifmmode {\cal L}\else ${\cal L}$\fi}
\def\inpb{\ifmmode {\rm pb}^{-1}\else ${\rm pb}^{-1}$\fi}
\def\infb{\ifmmode {\rm fb}^{-1}\else ${\rm fb}^{-1}$\fi}
\def\epem{\ifmmode e^+e^-\else $e^+e^-$\fi}
\def\ppb{\ifmmode \bar pp\else $\bar pp$\fi}
\def\mpl{\ifmmode \overline M_{Pl}\else $\overline M_{Pl}$\fi}
\def\cc{\ifmmode k/\overline M_{Pl}\else $k/\overline M_{Pl}$\fi}
\def\lpi{\ifmmode \Lambda_\pi\else $\Lambda_\pi$\fi}

\def\grtsim{\,\,\rlap{\raise 3pt\hbox{$>$}}{\lower 3pt\hbox{$\sim$}}\,\,}
\def\lsim{\,\,\rlap{\raise 3pt\hbox{$<$}}{\lower 3pt\hbox{$\sim$}}\,\,}
\newskip\zatskip \zatskip=0pt plus0pt minus0pt
\def\matth{\mathsurround=0pt}
\def\lsim{\mathrel{\mathpalette\atversim<}}
\def\gsim{\mathrel{\mathpalette\atversim>}}
\def\atversim#1#2{\lower0.7ex\vbox{\baselineskip\zatskip\lineskip\zatskip
  \lineskiplimit 0pt\ialign{$\matth#1\hfil##\hfil$\crcr#2\crcr\sim\crcr}}}

\def\be{\begin{equation}}
\def\ee{\end{equation}}
\def\bea{\begin{eqnarray}}
\def\eea{\end{eqnarray}}
\renewcommand{\thefootnote}{\fnsymbol{footnote}}

\begin{document} \begin{titlepage} 
\rightline{\vbox{\halign{&#\hfil\cr
&SLAC-PUB-8436\cr
&June 2000\cr}}}
\begin{center}

{\Large\bf  Experimental Probes of Localized Gravity:\\
On and Off the Wall}
\footnote{Work supported by the Department of 
Energy, Contract DE-AC03-76SF00515}
\medskip

\normalsize 
{\large H. Davoudiasl, J.L. Hewett, and T.G. Rizzo \\}
\vskip .3cm
Stanford Linear Accelerator Center \\
Stanford CA 94309, USA\\
\vskip .3cm

\end{center}

\begin{abstract} 

The phenomenology of the Randall-Sundrum model of localized gravity is 
analyzed in detail for the two scenarios where the Standard Model (SM) gauge 
and matter fields are either confined to a TeV scale 3-brane or may propagate 
in a slice of five dimensional anti-deSitter space.  In the latter instance, 
we derive the interactions of the graviton, gauge, and fermion Kaluza-Klein 
(KK) states.  The resulting phenomenological signatures are shown to be 
highly dependent on the value of the 5-dimensional fermion mass and differ 
substantially from the case where the SM fields lie on the TeV-brane.  In both 
scenarios, we examine the collider signatures for direct production of the 
graviton and gauge KK towers as well as their induced contributions to 
precision electroweak observables.  These direct and indirect signatures are 
found to play a complementary role in the exploration of the model parameter 
space.  In the case where the SM field content resides on the TeV-brane, we 
show that the LHC can probe the full parameter space and hence will either 
discover or exclude this model if the scale of electroweak physics on the 
3-brane is less than 10 TeV.  We also show that spontaneous electroweak 
symmetry breaking of the SM must take place on the TeV-brane.

\end{abstract}

\renewcommand{\thefootnote}{\arabic{footnote}} \end{titlepage} 


\section{Introduction}

A novel approach which exploits the geometry of extra spacetime dimensions 
has been recently proposed\cite{nima,rs,old} as a means 
to resolving the hierarchy problem.
In one such scenario due to Arkani-Hamed, Dimopoulos, and Dvali 
(ADD)\cite{nima}, the apparent hierarchy is generated by a large volume
for the extra dimensions.  In this case, the fundamental Planck scale in 
$4+n$-dimensions, $M$, can be brought down to a TeV and is related 
to the observed 4-d Planck scale through the volume $V_n$ of the compactified 
dimensions, $M^2_{Pl}=V_nM^{2+n}$.  In an alternative scenario due to
Randall and Sundrum (RS)\cite{rs}, the observed hierarchy is created by
an exponential warp factor which arises from a 5-dimensional
 non-factorizable geometry.  An exciting feature of these approaches is
that they both afford concrete and distinctive phenomenological 
tests\cite{pheno,dhr}.  Furthermore, if these theories truly describe the 
source of the observed hierarchy, then their signatures should appear in
experiment at the TeV scale.

The purpose of this paper is to explore the detailed phenomenology that
arises in the non-factorizable geometry of the RS model.  We will examine
the cases where the Standard Model (SM) gauge and matter fields can
propagate in the additional spacial dimension, denoted as the bulk, 
as well as being confined to
ordinary 3+1 dimensional spacetime.  The broad phenomenological features of
the latter case were spelled out in Ref. \cite{dhr}.  Here, we expand on this
previous work by considering the effects in precision electroweak observables
and investigating a wider range of collider signatures, including the case
of lighter graviton Kaluza-Klein (KK) excitations.  
We also show that the LHC can probe the full parameter space of this model
and hence will either discover or exclude it
if the scale of electroweak physics on the 3-brane is less than 10 TeV.
The experimental signatures of the former scenario, 
where the SM fields reside in the bulk, are considered here
for the first time.  As we will see below, this possibility introduces an
additional parameter, given by the 5-dimensional mass of the fermion fields, 
which has a dramatic influence on the phenomenological consequences and yields
a range of experimental characteristics.  While the general features of these
signatures remain indicative of this type of geometry, the various
details of the different cases
can be taken to represent a wide class of possible models similar
in nature to the RS scenario.  We also present an argument which shows
that spontaneous electroweak symmetry breaking must be confined to the 
Standard Model 3-brane.

The Randall-Sundrum model consists of a 5-dimensional non-factorizable
geometry based on a slice of AdS$_5$ space with length $\pi r_c$, where $r_c$
denotes the compactification radius.  Two 3-branes,
with equal and opposite tensions, rigidly reside at $S_1/Z_2$ orbifold
fixed points at the boundaries of the AdS$_5$ slice, taken to be 
$y=r_c\phi=0,r_c\pi$.  The 5-dimensional Einstein's
equations permit a solution which preserves 4-dimensional Poincar\' e
invariance with the metric
\be
ds^2=e^{-2\sigma(\phi)}\eta_{\mu\nu}dx^\mu dx^\nu - r_c^2d\phi^2\,,
\ee
where the Greek indices extend over ordinary 4-d spacetime and
$\sigma(\phi)=kr_c|\phi|$.  Here $k$ is the AdS$_5$ curvature scale which is of
order the Planck scale and is determined by the bulk cosmological constant
$\Lambda=-24M_5^3k^2$, where $M_5$ is the 5-dimensional Planck scale.  
The 5-d curvature scalar is then given by $R_5=-20k^2$.
Examination of the action in the 4-d effective theory yields the relation
\be
\mpl^2={M_5^3\over k}(1-e^{-2kr_c\pi})
\ee
for the reduced 4-d Planck scale.  
The scale of physical phenomena as realized by the 4-d flat metric 
transverse to the 5th dimension $y=r_c\phi$ 
is specified by the exponential warp factor.
TeV scales can naturally be attained on the 3-brane at $\phi=\pi$ if
gravity is localized on the Planck brane at $\phi=0$ and $kr_c\simeq 11-12$.
The scale of physical processes on this TeV-brane is then 
$\Lambda_\pi\equiv\mpl e^{-kr_c\pi}$.
The observed hierarchy is thus generated by a geometrical exponential
factor and no other additional large hierarchies appear.  It has been
demonstrated\cite{gw2} that this value of $kr_c$ can be stabilized without
the fine tuning of parameters by minimizing the potential for the modulus
field, or radion, which describes the relative motion of the 2 branes.
In the original construction of the RS model utilizing this stabilization 
mechanism, gravity and the modulus stabilization field may propagate freely
throughout the bulk, while the SM fields are assumed to be confined
to the TeV (or SM) brane at $\phi=\pi$.  The 4-d phenomenology of this model
is governed by only two parameters\cite{dhr}, given by the 
curvature $k$ and \lpi.  The radion, which receives a mass during the
stabilization procedure, is expected to be the
lightest new state and admits an interesting 
phenomenology\cite{radion} which we will not consider here.

This scenario has enjoyed immense popularity in the recent literature, with
the cosmological/astrophysical \cite{cosmo}, string 
theoretic\cite{stringy}, and phenomenological implications all
being explored.  We note that similar geometrical configurations have 
previously been found to arise in M/string theory\cite{strings}.  
In addition, extensions of this scenario where the higher dimensional 
space is non-compact\cite{rs2}, \ie, $r_c\to\infty$, 
as well as the inclusion of additional spacetime dimensions and 
branes\cite{morestuff} have been discussed.

Given the success of the RS scenario, it is logical to ask if it can be
extended to include other fields in the bulk besides gravity and the modulus
stabilization field.  It would appear to be more natural for all fields to
have the same status and be allowed to propagate throughout the full
dimensional spacetime.  
In addition, Garriga \etal\cite{spain} have recently
shown that the Casimir force of bulk matter fields themselves may be 
able to stabilize the radion field.  In the case of non-warped, toroidal
compactification of extra dimensions, bulk gauge fields can
lead to an exciting phenomenology which is accessible at 
colliders\cite{sminbulk,getV}.  The possibility of placing gauge fields in
the bulk of the RS model was first considered in Ref. \cite{dhr2}.  
In this case the couplings of the KK gauge bosons are greatly enhanced
in comparison to those of the SM by a factor of $\sqrt{2\pi kr_c}\simeq
8.4$.  An analysis of their contributions to electroweak radiative corrections
was found to constrain the mass of the first KK gauge boson excitation to
be in excess of 25 TeV, implying that the physical scale of the $\phi=\pi$
brane, $\Lambda_\pi$, must exceed 100 TeV.  By itself, if the model is to be
relevant to the hierarchy problem with $\Lambda_\pi$ being near the weak
scale, this disfavors the presence of SM gauge fields alone in the RS bulk.

This endeavor has recently been extended to consider fermion bulk fields.
Grossman and Neubert\cite{yuval} investigated this possibility 
in an effort to understand the neutrino mass hierarchy.  Using their results, 
Kitano\cite{kit} demonstrated that bounds on flavor changing
processes such as $\mu\to e\gamma$ also force the KK gauge bosons to be
heavy for neutrino Yukawa couplings of order unity.
Subsequently, Chang \etal\cite{chang} demonstrated that placing fermion fields
in the bulk allowed the zero-mode fermions, which are 
identified with the SM matter fields,
to have somewhat reduced couplings to KK gauge fields.  This
allows for a weaker constraint on the
value of \lpi\ from precision electroweak 
data.  Gherghetta and Pomarol\cite{gp} have noted the importance
of the value of the bulk fermion mass in determining the zero-mode fermion
couplings to both bulk gauge and wall Higgs fields and found interesting
implications for the fermion mass hierarchy and supersymmetry breaking.

In this paper we expand upon these studies and examine the phenomenological
implications of placing the SM gauge and matter fields in the bulk.  (In all 
cases to be discussed below, the backreaction on the metric due to the new 
bulk fields will be neglected.)   We find that this possibility introduces 
an additional parameter, given by the 5-dimensional
fermion mass, which governs the phenomenology.  In
the next section we peel the SM field content off  the TeV-brane, or wall,
and derive the KK spectrum and couplings of gravitons, bulk gauge fields,
and bulk fermions.  The 5-d fermion mass dependence of the 
couplings of the KK states to the zero-mode fermions is
explicitly demonstrated.  In section 3, we explore the phenomenology
associated with allowing the SM fields to propagate in the additional
dimension.  We delineate the broad
phenomenological features as a function of the bulk fermion mass and find
that there are four distinct classes of collider signatures.  We 
investigate these signatures and also compute the KK gauge contributions 
to electroweak radiative corrections.  We find that the stringent precision 
electroweak bounds on \lpi\ discussed above are significantly relaxed 
for a sizable range of the fermion bulk mass parameter.  In section 4,
we expand on our previous work\cite{dhr} and examine the phenomenology in
detail for the scenario where the SM fields all reside on the TeV-brane.  
Section 5
consists of our conclusions.  Appendix A contains an independent argument for
confining the Higgs fields to the TeV-brane.  Lastly, simplified 
expressions for a number of couplings as a function of the 
fermion bulk mass are given in Appendix B for the case when the SM field 
content propagates in the bulk.

\section{Peeling the Standard Model off the Wall}

In order to examine the phenomenological implications of placing the field 
content of the SM in the bulk of the RS model, we need
to know the properties of various bulk fields.  In this section, we 
review the KK reduction and interactions of massless  gravitons and bulk
gauge fields, as well as bulk fermions with arbitrary 5-d masses, and 
establish the notation that will be used in the sections
that follow.  Throughout our discussion, we will assume 
that the Higgs field and hence, spontaneous electroweak symmetry breaking, 
resides only on the TeV-brane.  This choice
has been advocated for a variety of different reasons 
by various authors{\cite {kit,chang,gp,higgsonbrane}}, and we will 
present an independent argument in Appendix A for keeping
the Higgs field on the TeV-brane.  We start our review with the 
massless bulk sector, namely the graviton and the gauge fields. 
In what follows, the Greek indices extend over the usual 4-d spacetime, 
whereas the upper case Roman indices represent all 5
dimensions.  The lower case Roman indices correspond to 
5-d Minkowski space. 

\subsection{Gravitons and Bulk Gauge Fields}

We parameterize the 5-d graviton tensor fluctuations 
$h_{\alpha \beta}$ ($\alpha, \beta = 0, 1, 2, 3$) by
\begin{equation}
\hat{G}_{\alpha \beta} = e^{-2 \sigma} \left(\eta_{\alpha \beta} + \kappa_5\,  
h_{\alpha \beta}\right),
\label{Ghat}
\end{equation}
where $\kappa_5 = 2 M_5^{-3/2}$ and the metric tensor is defined as
$\eta_{\mu\nu}={\rm diag}(1,-1,-1,-1)$.  
The 5-d graviton field $h_{\alpha \beta}(x, \phi)$ 
can be written in terms of a KK expansion of the form 
\begin{equation}
h_{\alpha \beta} (x, \phi) = \sum_{n = 0}^\infty h_{\alpha \beta}^{(n)} (x) \, 
\frac{\chi^{(n)}_G(\phi)}{\sqrt{r_c}}, 
\label{hKK}
\end{equation} 
where $h_{\alpha \beta}^{(n)}(x)$ represent the KK modes of the graviton 
(which we denote as $G^{(n)}$ in what follows)
with masses $m^G_n$ in 4-d Minkowski space and
$\chi^{(n)}_G(\phi)$ are the corresponding wavefunctions that 
depend only on the coordinate $\phi$ of the extra dimension.   

Employing the gauge choice $\eta^{\alpha \beta} \partial_\alpha 
h^{(n)}_ {\beta \gamma} = 0$ and $ \eta^{\alpha \beta} h_{\alpha
\beta}^{(n)} = 0$, and demanding the orthonormality condition 
\begin{equation}
\int_{- \pi}^{\pi} d \phi \, \, e^{- 2 \sigma} \chi^{(m)}_G \chi^{(n)}_G = 
\delta^{mn}, 
\label{horhto}
\end{equation}
we obtain \cite{rs,dhr}
\begin{equation}
\chi^{(n)}_G(\phi) = \frac{e^{2 \sigma}}{N_n^G} \left[J_2 (z_n^G) + 
\alpha_n^G \, Y_2 (
z_n^G)\right], 
\label{chih}
\end{equation}
where $J_q$ and $Y_q$ denote Bessel functions of order $q$ throughout 
this paper, $N_n^G$ give the 
wavefunction normalization, $\alpha_n^G$ are constant coefficients, and 
\begin{equation}
z_n^G(\phi) = m^G_n  \, \frac{e^{\sigma (\phi)}}{k}.
\label{zh}
\end{equation}
The solutions $\chi^{(n)}_G(\phi)$ are chosen to be $Z_2$-even in order to
obtain a massless zero-mode graviton.   The 
requirement of continuity of their first derivative at the orbifold fixed 
points $\phi = 0$ and $\phi = \pm \pi$ yields
\begin{equation}
\alpha_n^G \sim \left({x_n^G}\right)^2 e^{- 2 k r_c \pi}    
\label{alnh}
\end{equation}
and 
\begin{equation}
J_1 (x_n^G) = 0,
\label{xnh}
\end{equation}
where $x_n^G \equiv z_n^G(\phi = \pi)$, and we have assumed that 
$m^G_n/k \ll 1$ as well as $e^{k r_c \pi} \gg 1$.  
With these assumptions, we find $m^G_n = x_n^G\,  k \, e^{- k r_c \pi}$ and 
\begin{equation}
N_n^G \simeq \frac{e^{k r_c \pi}}{\sqrt{k r_c}} \, J_2 (x_n^G)\,; \quad \quad
\quad\quad n > 0.
\label{Nnh}
\end{equation}
The corresponding zero-mode is given by $\chi^{(0)}_G = \sqrt{k r_c}$.  We find 
$\alpha_n^G \ll 1$ for the KK 
modes of phenomenological importance, \ie, the lowest lying states, 
and thus the $Y_2$ term in Eq. (\ref{chih}) can be safely 
ignored compared to $J_2$ in our following analysis.  Note that the
masses of the graviton KK excitations are not equally spaced, unlike the case
for a factorizable geometry, with their separation here being dependent on the 
roots of $J_1$.  The first few values of $x_n^G$ are 3.83, 7.02, 10.17, and
13.32.

Next, we consider the case of a massless 5-d gauge field $A_M(x, \phi)$.  
Our notation is similar to that employed for the case of
the graviton field.  With the gauge choice $A_4(x, \phi) = 0$, and 
assuming that the KK expansion of $A_\mu(x, \phi)$ is given by 
\begin{equation} 
A_{\mu} (x, \phi) = \sum_{n = 0}^\infty A_{\mu}^{(n)} (x) \, 
\frac{\chi^{(n)}_A(\phi)}{\sqrt{r_c}}\,,
\label{AKK}
\end{equation}
the solutions for $\chi^{(n)}_A(\phi)$ are \cite{dhr2}
\begin{equation} 
\chi^{(n)}_A = \frac{e^{\sigma}}{N_n^A} \left[J_1 (z_n^A) + 
\alpha_n^A \, Y_1 (z_n^A)\right]\,,
\label{chiA}
\end{equation}
subject to the orthonormality condition 
\begin{equation} 
\int_{- \pi}^{\pi} d \phi \, \chi^{(m)}_A \chi^{(n)}_A = \delta^{mn}.
\label{Aortho}
\end{equation}
The functions $\chi^{(n)}_A$ in Eq. (\ref{chiA}) are also chosen 
to be $Z_2$-even.  The continuity of $d \chi^{(n)}_A/d\phi$ at
$\phi = 0$ yields
\begin{equation} 
\alpha_n^A = - \frac{J_1 (m_n^A/k) + (m_n^A/k)  J_1^\prime (m_n^A/k)}
{Y_1 (m_n^A/k) + (m_n^A/k)  Y_1^\prime (m_n^A/k)}\,,
\label{alAn}
\end{equation}
and at $\phi = \pm \pi$ we obtain
\begin{equation}  
J_1 (x_n^A) + x_n^A  J_1^\prime (x_n^A) + \alpha_n^A 
\left[Y_1 (x_n^A) + x_n^A  Y_1^\prime (x_n^A)\right] = 0,
\label{xAneq}
\end{equation}
where $m_n^A$ is the mass of the $n$th KK mode of the 
gauge field with $m_n^A = x_n^A \, k e^{- k r_c \pi}$.  Again, we see that
the masses of the gauge KK excitations are not equally spaced.  The 
normalization $N_n^A$ is given by \cite{chang}
\begin{equation} 
N_n^A = \left(\frac{e^{k r_c \pi}}{x_n^A \, \sqrt{k r_c}}\right) 
\sqrt{\left\{{z_n^A}^2 \left[J_1 (z_n^A) + 
\alpha_n^A \, Y_1 (z_n^A)\right]^2\right\}^{z_n^A(\phi = 
\pi)}_{z_n^A(\phi = 0)}} \, \, . 
\label{NAn}
\end{equation} 
The zero-mode gauge field is then $\chi^{(0)}_A = 1/\sqrt{2 \pi}$. 
The first few numerical values of $x_n^A$ are 2.45, 5.57, 8.70, and 11.84.

\subsection{Bulk Fermion Fields}

We now discuss the KK solutions for bulk fermions \cite{yuval,kit,chang,gp} 
of arbitrary Dirac 5-d mass; the possibility of  Majorana mass terms for 
neutral fermion fields will not be considered here.  
The action $S_f$ for a 
free fermion of mass $m$ in the 5-d RS model is \cite{yuval}
\begin{equation}
S_f =\int d^4x \int d\phi \sqrt{G} \left[V^M_n\left
({i\over {2}}\overline{\Psi} \, \gamma^n\,  \partial_M \Psi + h. c.\right) - 
sgn(\phi) m \overline{\Psi} \Psi\right],
\label{Sf}
\end{equation}
where $h. c.$ denotes the Hermitian conjugate term, and we have 
$\sqrt{G} = [det (G^{MN})]^{1/2}=e^{-4\sigma}$, $n = 0, 1, \ldots, 4$,
$V^M_\mu = e^\sigma \delta^M_\mu$, $V^4_4 = -1$, and 
$\gamma^n = (\gamma^\nu, i \gamma_5)$. As demonstrated  
previously{\cite {yuval,chang,gp}}, the contribution to the 
action from the spin 
connection vanishes when the hermitian conjugate term is included.
The form of the mass term
is dictated  by the requirement of $Z_2$-symmetry \cite{yuval} since 
$\overline \Psi \Psi$ is necessarily odd under $Z_2$ as can be seen
from examining the first term in the action.
We adopt the notation of Ref. \cite{yuval} for the KK expansion of
the $\Psi$ field and write 
\begin{equation} 
\Psi_{L, R}(x, \phi) = \sum_{n = 0}^\infty \psi_{L, R}^{(n)}(x) \, 
\frac{e^{2 \sigma(\phi)}}{\sqrt{r_c}}\, \hat{f}^{(n)}_{L, R}(\phi), 
\label{PsiKK} 
\end{equation}
where $L$ and $R$ refer to the chirality of the fields and $\hat f_{L,R}^{(n)}$
represent 2 distinct complete orthonormal functions.
The orthonormality relations are then given by
\begin{equation}
\int_{-\pi}^\pi d \phi \, \, e^\sigma {\hat{f}^{(m) *}_{L}} 
\hat{f}^{(n)}_{L} = 
\int_{-\pi}^\pi d \phi \, \, e^\sigma 
{\hat{f}^{(m) *}_{R}} \hat{f}^{(n)}_{R} = \delta^{mn}.    
\label{fortho} 
\end{equation}
Due to  the requirement of $Z_2$-symmetry of the action, 
$\hat{f}^{(n)}_{L}$ and $\hat{f}^{(n)}_{R}$ must have opposite 
$Z_2$-parity;  here we choose $\hat{f}^{(n)}_{L}$ to be 
$Z_2$-even and $\hat{f}^{(n)}_{R}$ to be $Z_2$-odd.  The  
SM matter fields then correspond to  the zero-modes  
 $\hat{f}^{(0)}_{L}$.  
All of the SM fermion fields are thus treated as left-handed as
is commonly done in the literature. 
The KK reduction of the 
action $S_f$ through the expansion 
(\ref{PsiKK}) for $\Psi_{L, R}(x, \phi)$ yields the solutions 
\begin{equation}    
\hat{f}^{(n)}_{L, R}(\phi) = \frac{e^{\sigma/2}}{N_n^{L, R}} 
\left[J_{\frac{1}{2} \mp \nu}(z_n^{L, R}) + \beta_n^{L, R} \,
Y_{\frac{1}{2} \mp \nu}(z_n^{L, R})\right]
\label{fLR} 
\end{equation}
for $n \neq 0$.  The zero-mode $\hat{f}^{(0)}_{L}$, corresponding 
to a massless 4-d SM fermion, is given by  
\begin{equation}
\hat{f}^{(0)}_{L} = \frac{e^{\nu \sigma}}{N_0^L}.
\label{f0} 
\end{equation}
Here $\nu$ is defined by $m\equiv\nu k$ and is expected to be of order
unity.  For simplicity and phenomenological reasons
we take all fermions to have the same value of
$\nu$ throughout this paper.

With our choices for the $Z_2$-parity of the wavefunctions, 
the coefficients $\beta_n^{L, R}$ and the masses 
$m_n^{L, R}$ of the KK modes are obtained by requiring 
\begin{equation}
\left(\frac{d}{d \phi} - m \, r_c\right) \hat{f}^{(n)}_{L} = 0
\label{condL} 
\end{equation}
and 
\begin{equation}    
\hat{f}^{(n)}_{R} = 0 
\label{condR} 
\end{equation}
at $\phi = 0, \pm \pi$, for the left- and right-handed solutions, 
respectively.  In the case of the left-handed wavefunctions, 
we obtain
\begin{equation}
\beta_n^{L} = - \frac{J_{- (\nu + \frac{1}{2})}(m_n^L/k)}{Y_{- (\nu + 
\frac{1}{2})}(m_n^L/k)}
\label{betaL} 
\end{equation} 
from evaluating the above conditions at $\phi = 0$, and 
\begin{equation}
J_{- (\nu + \frac{1}{2})}(x_n^L) + \beta_n^{L} \,  
Y_{- (\nu + \frac{1}{2})}(x_n^L) = 0
\label{Lroots} 
\end{equation}
at $\phi = \pi$.
Similarly, for the right-handed solutions, we have 
\begin{equation}
\beta_n^{R} = - \frac{J_{\nu + \frac{1}{2}}(m_n^R/k)}{Y_{\nu + 
\frac{1}{2}}(m_n^R/k)}
\label{betaR} 
\end{equation} 
 and
\begin{equation}   
J_{\nu + \frac{1}{2}}(x_n^R) + 
\beta_n^{R} \,  Y_{\nu + \frac{1}{2}}(x_n^R) = 0\,.
\label{Rroots} 
\end{equation}
Note that the left- and right-handed excitation masses, $m_n^{L,R}$, 
are degenerate for each value of $n$ above the 
zero-mode. The orthonormality of $\hat{f}^{(n)}_{L, R}$ yields
\begin{equation}   
N_0^L = \sqrt{\frac{2\left[e^{k r_c \pi (1 + 2 \nu)} - 
1\right]}{k r_c (1 + 2 \nu)}}
\label{Nf0} 
\end{equation}
and
\begin{equation} 
N_n^{L, R} = \left(\frac{e^{k r_c \pi}}{x_n^{L, R} \, \sqrt{k r_c}}\right) 
\sqrt{\left\{{z_n^{L, R}}^2 \left[J_{\frac{1}{2} \mp \nu}(z_n^{L, R}) + 
\beta_n^{L, R} \, Y_{\frac{1}{2} \mp \nu}(z_n^{L, R})\right]^2
\right\}^{z_n^{L, R}(\phi = \pi)}_{z_n^{L, R}(\phi = 0)}} \, \, . 
\label{Nfn}
\end{equation}

We note here that only the left-handed fermion fields are relevant
to the phenomenological study in this paper, 
since their zero-modes correspond to the SM fermions.

\nn
\begin{figure}[htbp]
\centerline{
\psfig{figure=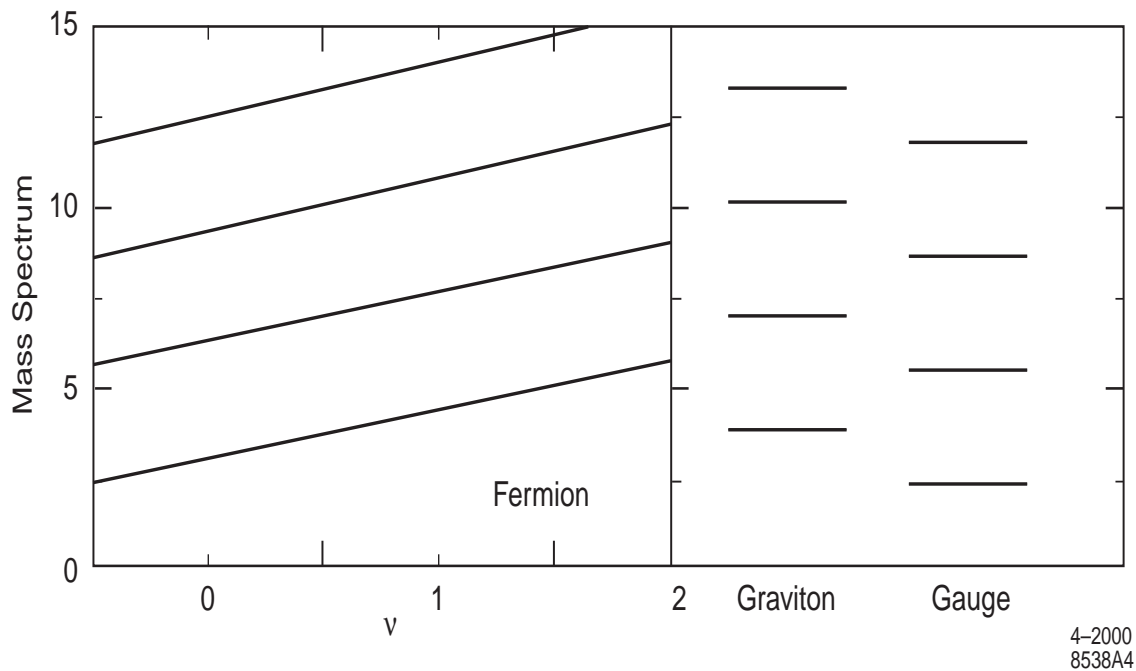,height=9cm,width=15cm,angle=0}}
\caption{Relative mass spectra in units of $ke^{-kr_c\pi}$ of the KK 
excitations of the 
fermion fields as a function of their bulk mass parameter $\nu$, as well as
for the graviton and the gauge boson fields as described in the text.}
\label{fig1}
\end{figure}

Given the above set of equations we can determine the relative 
values for the masses of the KK 
states for the graviton, gauge, and fermion tower members by numerically 
solving for the appropriate Bessel function roots.  Recall that 
degenerate right- and left-handed fermion KK towers both exist 
for the fermion states that lie above the left-handed zero-modes.
These mass spectra are displayed in 
Fig. \ref{fig1} in units of 
$ke^{-kr_c\pi}$.  The fermion KK excitation masses have an approximately linear 
dependence on $\nu$ given by
$m_n^f\simeq a_n|\nu+1/2|+b_n$, with $a_n,b_n$ being essentially constant 
for each tower member.  For the values $\nu <-1/2$, 
we find that the fermion masses are simply reflected about the point
$\nu=-1/2$, with $m_n^f(\nu)=m_n^f(-[\nu+1])$, implying that the lightest 
fermion KK states occur when $\nu=-1/2$.  
Note that at $\nu=-1/2\, (+1/2)$ fermions and gauge 
bosons (gravitons) are predicted to be degenerate in mass.  In addition,
the fermion excited KK states
are generally expected to be more massive than the corresponding gauge
boson states.

\subsection{Couplings of the KK Modes}

Having reviewed the KK reduction of various SM bulk fields in the 
RS model, we now turn our attention to the couplings of 
the KK modes in the 4-d effective theory.  We focus on the
vertices that are of relevance to the phenomenology 
discussed in this work.  In what follows, we give the 
integrals that yield the couplings of fermions to gravitons and
gauge fields and evaluate their dependence on the fermion bulk mass
in the case where the SM matter fields propagate in the bulk.
In addition, we provide the coupling of gauge fields to gravitons and
discuss the interactions between zero-mode fermion and gauge KK
states with a Higgs field confined to the TeV-brane.
In Appendix B, we present simplified expressions for 
these integrals as well as for a number of additional 3-point functions. 

Schematically, the coupling of the $m$th and $n$th KK modes of the field 
$F$ to the $q$th KK level graviton is given by
\begin{equation}
S_G =  \sum_{m,n,q} \left\{\left[\int \frac{d\phi}{\sqrt{k}} \, 
\frac{e^{t \sigma} \, \, \chi_{F}^{(m)} 
\chi_{F}^{(n)}  
\chi_G^{(q)}}{\sqrt{r_c}}\right] \frac{\kappa_4}{2}\int d^4x \, 
\eta^{\mu \alpha} \, \eta^{\nu \beta} h^{(q)}_{\alpha \beta}(x) \, 
T^{(m,n)}_{\mu \nu}\right\},     
\label{SG}
\end{equation}
where $t$ depends on the type of field $F$,
$\chi_{F}^{(n)}$ represents the $n$th 
KK solution of the field $F$, 
$\chi_G^{(q)}$ is the $q$th KK graviton wavefunction, 
$h^{(q)}_{\alpha \beta}(x)$ corresponds to the $q$th KK graviton mode, 
$\kappa_4/2 = \overline{M}_{Pl}^{- 1}$, and 
$T^{(m,n)}_{\mu \nu}$ denotes the 4-d energy 
momentum tensor for the fields.  The information regarding
the spacetime curvature and the shape of the wavefunctions in 
the 5th dimension is encoded in a coefficient 
$C$ given by the integral in brackets above,
\begin{equation}
C_{mnq}^{FFG} = \int \frac{d\phi}{\sqrt{k}} \, \frac{e^{t \sigma} 
\, \, \chi_{F}^{(m)} 
\chi_{F}^{(n)} \chi_G^{(q)}}{\sqrt{r_c}}\,.
\label{C}
\end{equation}
To compute the coupling of $F$ to a KK graviton in the RS model, 
one must multiply the corresponding Feynman rules derived in
flat spacetime with extra dimensions \cite{pheno}, which are
written in terms of  $T^{(m,n)}_{\mu \nu}$,  by
$C_{mnq}^{FFG}$.  We now present these coefficients for 
the cases of fermion and gauge field interactions with the KK
graviton states.   Note that with the conventions discussed above for the 
wavefunctions of various bulk fields, the coupling strength
of the zero-mode graviton is fixed to be $\mpl^{- 1}$ 
in the 4-d effective theory.

For the case where the SM fields propagate in the bulk,
the coefficient $C^{f\bar fG}_{m n q}$ of the coupling of the 
$m$th and the $n$th fermion KK states to the $q$th 
graviton mode can be obtained from the term
\begin{equation}
S_1 = i \int d^5x \, \sqrt{G} \, \, V^M_n \, \overline{\Psi} \, \gamma^n\,  
\partial_M \Psi 
\label{S1}
\end{equation}
in the action, and is given by
\begin{equation}     
C^{f\bar fG}_{m n q} = \int_{-\pi}^\pi \frac{d\phi}{\sqrt{k}} \, 
\frac{e^\sigma \hat{f}^{(m)}_L \hat{f}^{(n)}_L \chi^{(q)}_G}{\sqrt{r_c}}.
\label{Cfmnp}
\end{equation}
The corresponding coefficient $C^{AAG}_{m n q}$ for the coupling strength of 
the $m$th and the $n$th KK excitations of a gauge field to the $q$th 
graviton mode, can be deduced from the interaction
\begin{equation}
S_2 = {-1\over 4} \int d^5x \, \sqrt{G} \, G^{MA} G^{NB} F_{AB} F_{MN}\,,
\label{S2}
\end{equation}
yielding
\begin{equation}  
C^{AAG}_{m n q} = \int_{-\pi}^\pi \frac{d\phi}{\sqrt{k}} \, \frac{\chi^{(m)}_A 
\chi^{(n)}_A \chi^{(q)}_G}{\sqrt{r_c}}.
\label{CAmnp}
\end{equation}

Next, we consider the interaction between a fermion field $\Psi$ and a
gauge field $A_M$.  The coefficient of this coupling 
is obtained from the interaction
\begin{equation} 
S_3 = \int d^5x \, \sqrt{G} \, \, V^M_n \, g_5 \, \overline{\Psi} \, 
\gamma^n \,  A_M \Psi, 
\label{S3}
\end{equation}
where $g_5$ is the 5-d gauge coupling constant.  
Since the zero-mode wavefunction for the field $A_\mu (x, \phi)$ is given by 
$\chi^{(0)}_A = 1/\sqrt{2 \pi}$, the interaction 
of zero-mode fermion and gauge fields is given by
\begin{equation} 
S_3 = \frac{g_5}{\sqrt{2 \pi r_c}} \int d^4x \, \eta^{\mu \nu} \, 
\overline{\psi}^{(0)} \, 
\gamma_\mu \,  \psi^{(0)} A_\nu^{(0)} + \ldots,
\label{S3000}
\end{equation}
where we have used the orthonormality of the fermion wavefunctions 
given by Eq. (\ref{fortho}).  We thus see that $g_4 = 
g_5/\sqrt{2 \pi r_c}$, where $g_4$ is the usual 4-d SM gauge coupling.  
In general, the coefficient $C^{f\bar fA}_{m n q}$ of 
the coupling of the $m$th and the $n$th fermion states to the 
$q$th gauge field mode, in units of $g_4$, is given by
\begin{equation} 
C^{f\bar fA}_{m n q} = \sqrt{2 \pi} \int_{-\pi}^\pi d\phi \, e^\sigma 
\hat{f}^{(m)}_L \hat{f}^{(n)}_L \chi^{(q)}_A.  
\label{DfAmnp}
\end{equation} 
With these general expressions it is straight-forward to compute the 
couplings of  any number of 
gauge, fermion, and graviton fields. In Appendix B we provide a set of 
useful couplings expressed in  simplified form.

\begin{figure}[t]
\centerline{
\psfig{figure=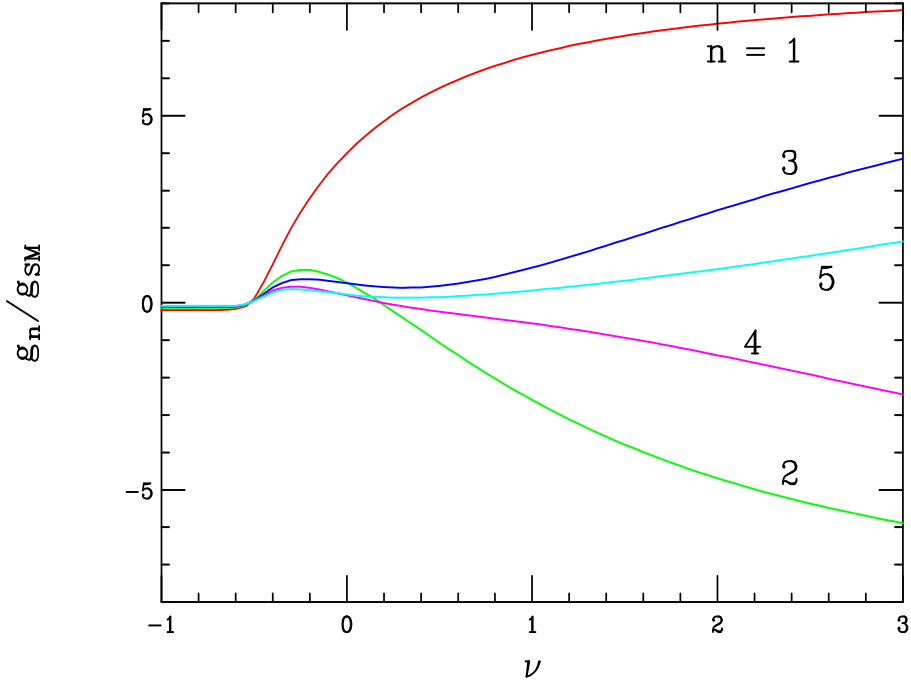,height=9cm,width=12cm,angle=90}}
\caption{The coupling strength of the zero-mode fermions to the first five KK 
gauge boson states in units of the corresponding SM coupling strength 
as a function of $\nu$. 
From top to bottom on the right-hand side of the figure the curves are for 
the first, third, fifth, fourth and second gauge KK excitations.}
\label{fig2}
\end{figure}

\nn
\begin{figure}[htbp]
\centerline{
\psfig{figure=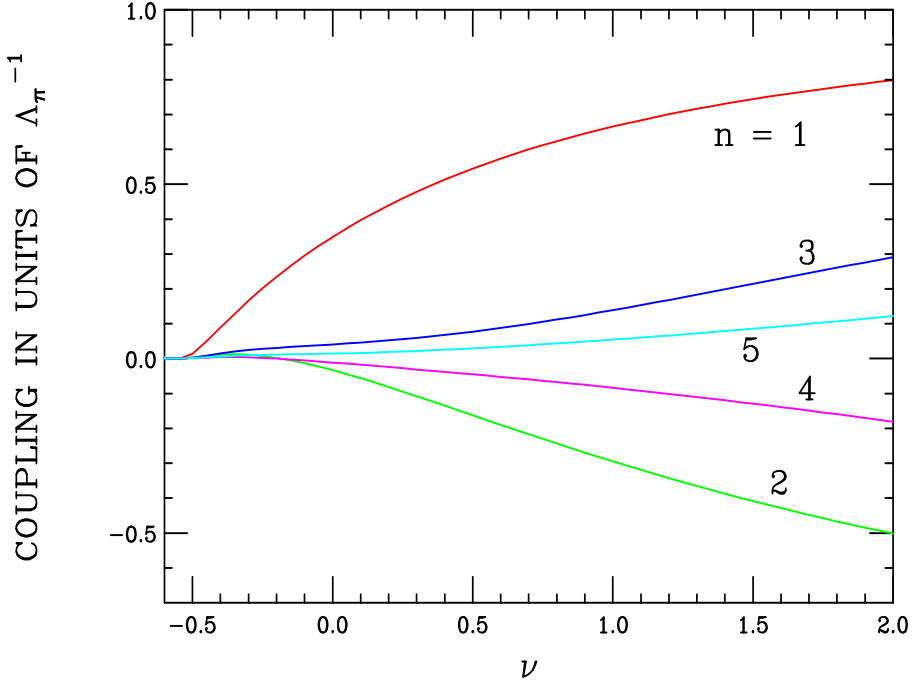,height=9cm,width=12cm,angle=90}}
\caption{The coupling strength of the zero-mode fermions to the first five KK 
graviton states in units of $\Lambda_\pi^{-1}$ as a function of $\nu$. 
From top to bottom on the right-hand side of the figure the curves are for 
the first, third, fifth, fourth and second graviton KK levels.}
\label{fig3}
\end{figure}

For the practical applications considered in this paper we need to determine
the detailed dependence on $\nu$ of the couplings of 
the zero-mode fermions to the members of the gauge and graviton KK towers, as 
well as the couplings of the zero-mode gauge fields to the graviton tower. 
Simplified versions of these specific couplings can be found in Appendix B 
in Eqs. (51-53). Figure \ref{fig2} displays
the couplings of the zero-mode fermions to the gauge KK tower members in 
units of the corresponding SM coupling strength. This result reproduces 
that of Ref.
{\cite {gp}} with their parameter $c$ being identified as $-\nu$. Note 
that as $\nu$ becomes large, which means that the fermion wavefunctions are
localized 
closer to the SM brane, the magnitude of the gauge couplings grow 
significantly. For $\nu \gg 1$ we recover the result for the case where
the SM fermions are confined to the TeV-brane, \ie, 
that $|g^{(n)}/g_{SM}| \to \sqrt {2\pi kr_c}$. On the other hand, for 
of $\nu\lsim -0.5$, the couplings become quite small and 
are approximately independent of $\nu$.  We then expect to
obtain strong direct and indirect bounds on the gauge KK states for
$\nu \gsim -0.3$, 
while for smaller values of $\nu$ there will be a serious degradation in the 
ability of experiment to probe large KK mass scales. Note that the gauge 
tower couplings essentially vanish in the region near $\nu=-0.5$. 

The corresponding $\nu$-dependent couplings of the graviton KK tower states 
to the zero-mode fermions are 
displayed in Fig. \ref{fig3}.  Here, we have taken the coefficient 
given by Eq. (52) in the Appendix and included the factor of 
$\kappa_4/2$ in Eq. (30) to
obtain the full coupling strength which is in units of $\lpi^{-1}$.
Again, as $\nu \gg 1$ the magnitude of the 
coupling strength for each tower member approaches unity in units of 
$\Lambda_\pi^{-1}$ which is the well-known result for wall fermions. 
However, for values of $\nu$ below $\nu \simeq -0.5$, the gravitational 
couplings of the zero-mode fermions become exponentially small for all 
massive graviton tower members, \ie, the fermions essentially decouple from 
the KK graviton states. This will make it 
impossible in this region to search, either directly or indirectly, for the
graviton KK excitations via their interactions with fermions. 

The couplings of zero-mode gauge fields to the graviton KK tower are, of 
course, independent of $\nu$ as can be seen from Eq. (53) in the Appendix. 
For the first five 
KK graviton tower members we find these couplings to be 1.34, 0.268, 0.273, 
0.114, and 0.127 in units of 
$10^{-2}\Lambda_\pi^{-1}$. Note that the strength of these couplings are all 
small, implying that searches for gravitons via these interactions will 
also be rather difficult. 

The couplings of the zero-mode fermion and gauge bulk fields to the Higgs 
when the Higgs is constrained to lie on the TeV-brane 
are also important since these are responsible for spontaneous symmetry 
breaking. These are also discussed in Appendix B. We find that 
in terms of a dimensionless Yukawa coupling in 5-d, $\tilde \lambda_5$, the 
corresponding 4-d Yukawa coupling for zero-mode fermions is given by
\begin{equation}
\lambda_4={\tilde \lambda_5\over 2} 
\left[{1+2\nu\over {1-\epsilon^{1+2\nu}}}\right]\,,
\end{equation}
with $\epsilon\equiv e^{-kr_c\pi}$.  This reproduces the result of 
Ref. \cite{gp}.  Note that the function in the square bracket is continuous 
and equal to unity when $\nu=-1/2$.
If one assumes that $\tilde \lambda_5$ is of order unity, then we see 
that $\lambda_4$ is also of order unity provided $\nu \gsim -0.5$.  For 
smaller values of $\nu$ the magnitude of the 4-d Yukawa coupling
falls rapidly, \eg, if $\nu=-0.75$ then $\lambda_4 \sim \sqrt \epsilon 
\sim 10^{-8}$.  Even if one allowed for fine 
tuning, this implies that it would be difficult to generate the observed SM
fermion mass spectrum for values of $\nu \lsim  -0.8$ to $-0.9$.  We thus
restrict ourselves to the region $\nu\gsim -0.8$ in our phenomenological
discussions below. 
Similar arguments also show that the vacuum expectation value of the 
Higgs on the TeV-brane
naturally leads to the conventional masses for the $W$ and $Z$ gauge bosons
which we identify as the zero-mode members of their respective towers. 

\section{Phenomenology of Bulk Fields}

In comparison to the analyses of the RS model where the SM field content
is confined to the TeV-brane, the phenomenology for the case where both 
SM gauge fields and fermions 
are allowed to propagate in the bulk is more complex due to the
a priori unknown value of the bulk fermion mass parameter $\nu$. In 
what follows, for simplicity, and to avoid problems with proton decay 
and flavor changing neutral current effects{\cite {gp}}, we will 
assume that all SM fermions have 
the same value of $\nu$.  Here we employ a two-pronged attack on the 
model by examining its implications on both precision electroweak measurements 
and direct collider searches.  We will see that the two techniques provide 
complementary information and constraints, as is usually the case, with the
conclusion being that the range of $\nu$ over 
which the RS model with SM fields in the bulk provides a solution to the 
hierarchy problem without being overly 
fine-tuned, \ie, values of $\Lambda_\pi\lsim 10$ TeV, is a rather 
small fraction of what is allowed by naturalness arguments. 

\subsection{Precision Electroweak Observables}

As is well-known, precision electroweak data can be used to place 
complementary constraints on new physics scenarios to those obtainable from 
direct collider searches\cite{htt}.  The analysis we employ below is a 
natural extension to that developed earlier by Rizzo and Wells\cite{getV}
in the case of the 5-dimensional SM with a factorizable geometry with gauge
bosons alone being in the bulk.
In that work, a global analysis was performed of the KK gauge tower
tree-level contributions to a large set of 
electroweak observables: $M_W$, $Z$-boson partial widths and asymmetries,
$\sin^2\theta_w$, atomic parity violation expressed via the weak charge
$Q_w$\cite{apv}, and the Paschos-Wolfenstein\cite{mani} asymmetry $R^-$
as measured by the NuTeV/CCFR collaboration\cite{nutev}.  In this scenario,
the gauge KK states above the zero-mode are evenly spaced and all couple 
with the same strength, and the authors\cite{getV} concluded that the
mass of the lightest KK excitation of the SM gauge fields must be in
excess of 3.3 TeV.  This result is similar in magnitude
to the corresponding limits
obtainable from contact interaction analyses\cite{cints}.  This procedure
has also been employed\cite{dhr2} in the case where the gauge bosons are the
only SM fields to propagate in the non-factorizable RS bulk.  
In this case, the couplings of the KK
tower members to the wall fermions are also independent of the
particular KK state above the zero-mode, but the ratio of the fermionic
couplings of the $n$th excitation to those of the zero-mode
is large with $g_n/g_0=\sqrt{2\pi kr_c}\simeq
8.4$ and the masses of the tower members are no longer equally spaced,
being given by roots of the appropriate Bessel functions as discussed
above.  There it was\cite{dhr2} found that the first SM gauge KK excitation 
must be more massive than $\simeq 23$ TeV.

\begin{figure}[htbp]
\centerline{
\psfig{figure=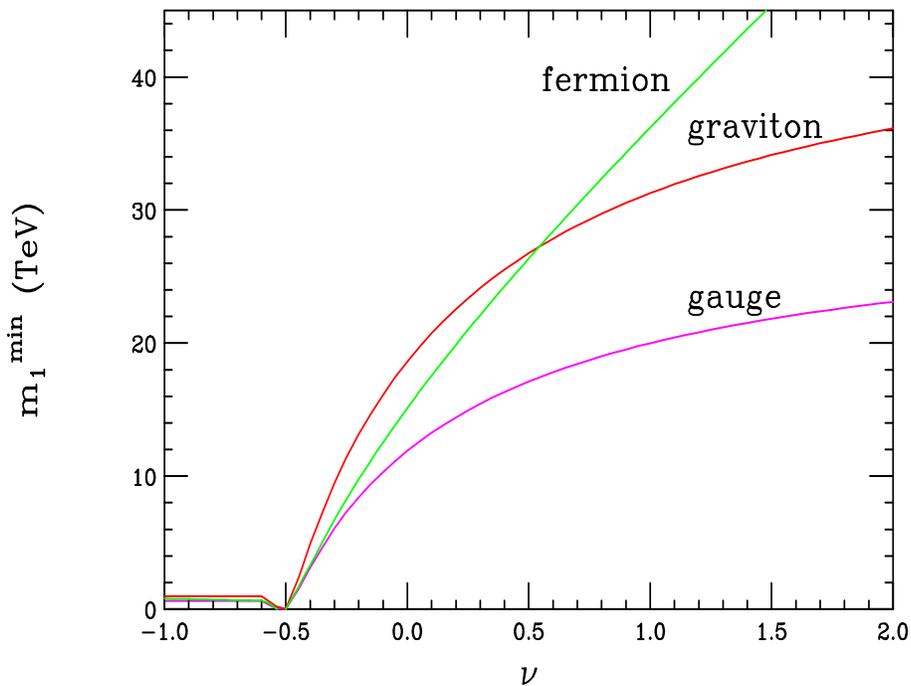,height=9cm,width=12cm,angle=90}}
\caption{The bounds on the masses of the lightest graviton, gauge boson and 
fermion KK state as a function of $\nu$ as obtained from the analysis of 
radiative corrections discussed in the text and the use of the mass 
relationships shown in Fig. \ref{fig1}. 
From top to bottom on the right-hand side the curves 
correspond to the mass of the lightest fermion, graviton and gauge KK states.}
\label{fig4}
\end{figure}

Here, the situation is more complex since once the fermions are allowed
to reside in the bulk, each member of the gauge KK tower couples to
the zero-mode fermions with a different strength, which is dependent
on the parameter $\nu$ as discussed above.  Following the analyses of
Ref. \cite{getV,dhr2}, we work in the limit where the KK tower exchanges
can be characterized as a set of contact interactions by integrating out the
tower fields.  The tower exchanges then lead to new dimension-six operators
whose coefficients are proportional to
\be
V(\nu)=\sum^\infty_{n=1} {g_n^2(\nu)\over g^2_0}{M_W^2\over m_n^2}\,,
\ee
where $g_n(\nu)$ is the $\nu$ dependent coupling of the $n$th tower member
with mass $m_n$, and $g_0$ is identified as the corresponding SM coupling.
The $g_n(\nu)$ for the gauge KK fields were computed in the previous section and
are given in Appendix B.  A global fit to the most recent electroweak data 
as presented at Moriond 2000\cite{moriond} for the observables listed above, 
results in somewhat stronger bounds on the quantity $V$ than those
obtained earlier\cite{getV,dhr2}, mainly due to the new value of
$Q_w$ \cite{apv} employed in the fit.  The resulting lower bound on
the mass of the first gauge KK state as a function of $\nu$ is shown
in Fig. \ref{fig4}.  Using the mass relationships given in the previous
section between the gauge, graviton, and fermion KK excitations, 
we can translate this bound into constraints on the masses of the 
other first tower members as well; this is also displayed in the figure.
Note that as $\nu$ becomes large and positive we reproduce the constraint
computed in Ref. \cite{dhr2} for the case where the fermions are on the
wall \ie, $m_1^{\rm gauge}\gsim 25$ TeV, which translates into the bound
$\Lambda_\pi \gsim 100$ TeV. However, for 
smaller values of $\nu$, values of $\Lambda_\pi$ of order a few TeV or less are 
clearly consistent with the data. The general $\nu$ dependent
behavior of these constraints can be easily understood from the values of
$g_n(\nu)/g_0$ shown in Fig. \ref{fig2}.  Recall that for $\nu\lsim -0.5$, 
the gauge tower couplings are small and approximately $\nu$ independent, 
while for $\nu\gsim -0.5$, the tower couplings grow rapidly with increasing 
values of $\nu$.  Hence, the precision electroweak bounds on the first tower
states are rather weak and $\nu$
independent with $m_1^{\rm gauge}\gsim 620$ GeV for $\nu\lsim -0.5$, and
disappear completely for $\nu=-0.5$, but grow
rapidly with increasing values of $\nu$ reaching the multi-TeV region.

\nn

While almost all of the observables used in the electroweak fit described 
above are $\nu$-dependent since fermion couplings are directly involved, one
is not, namely the mass of the $W$.  Hence, one might be tempted to 
obtain a $\nu$-independent bound by using just this quantity alone. 
Unfortunately, a useful limit cannot be obtained using this single observable 
without a priori knowledge of the Higgs boson mass. As was shown in the 
analysis of Rizzo and Wells{\cite {getV}} for Higgs fields on the wall, 
the existence of KK tower states for both the $W$ and $Z$ gauge fields 
will lead to a predicted increase in  $M_W$ for a fixed value of the 
Higgs mass when $M_Z$ is used as input. However, this increase in $M_W$ due 
to KK excitations can always be offset by a compensating increase in the 
Higgs mass which in turn
lowers $M_W$ due to loop effects.  Thus, unless the Higgs mass is otherwise
determined, one can 
always have a trade off between the gauge KK tree level and Higgs boson 
loop contributions.  Once the Higgs mass is 
known, however, a $\nu$-independent bound can be obtained. This point has 
recently been emphasized by Kane and Wells{\cite {gandj}}. We note that in 
performing the global fit described above, the only 
assumption about the Higgs mass was that $m_H\geq 100$ GeV.

\subsection{Collider Studies}

\begin{figure}[htbp]
\centerline{
\psfig{figure=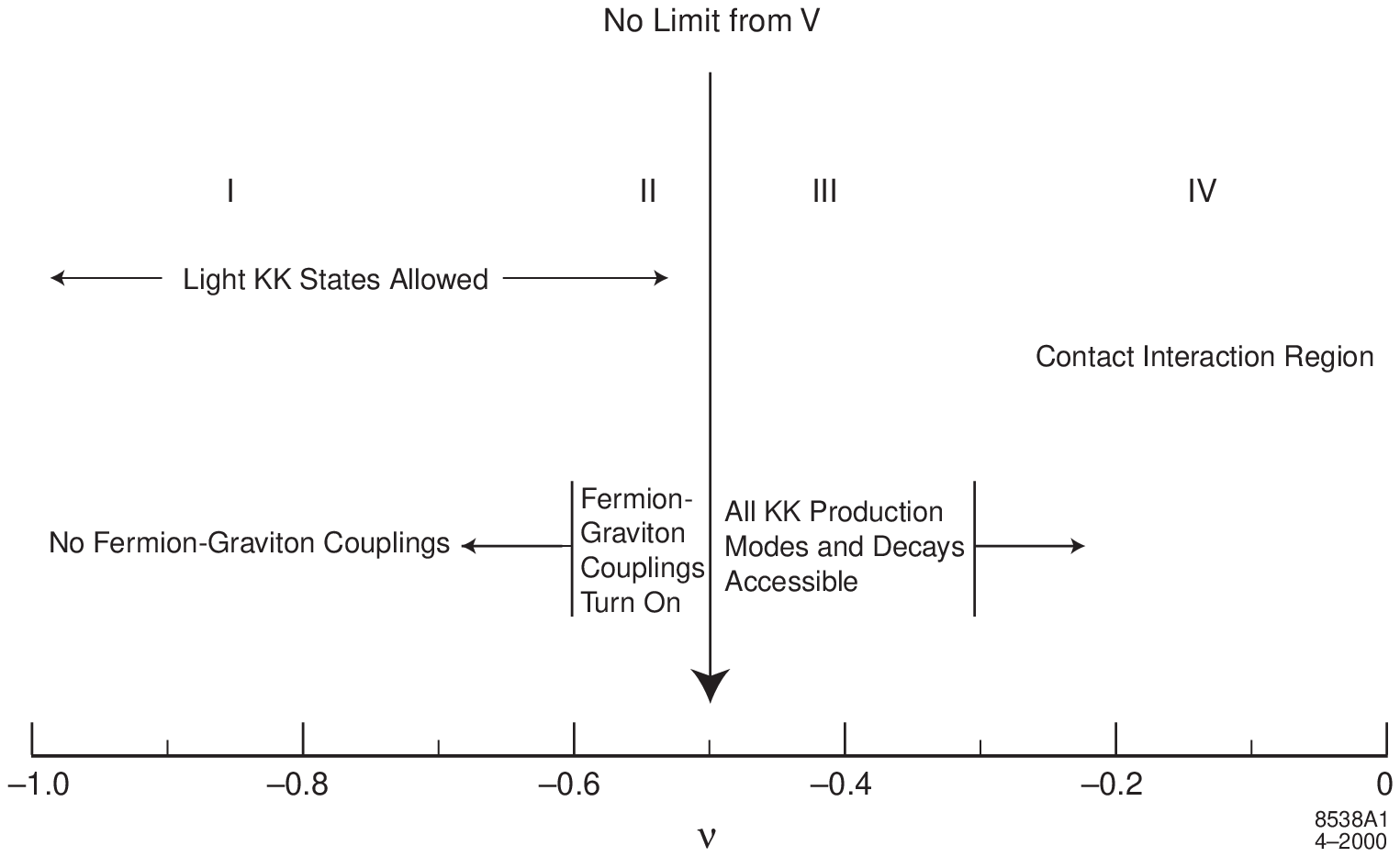,height=9cm,width=15cm,angle=0}
}
\caption{The descriptive phenomenology for each
region of $\nu$ as discussed in the text.}
\label{fig5}
\end{figure}

It is clear from the results shown in Figures 2, 3, and 4 that 
four distinct regions, corresponding to specific ranges of $\nu$, emerge,
yielding four different classes of phenomenology.
This is described in Fig. \ref{fig5}. Region I corresponds to the
range $ -0.9$ to $-0.8\lsim\nu\lsim -0.6$, where the lower boundary is set
by not allowing the fermion Yukawa couplings to be fine-tuned, as
discussed in the previous section.  Here, the SM fermions have decoupled 
from the graviton KK tower and are only very weakly coupled to the gauge KK 
states. (Recall that the SM gauge fields only interact weakly with to the 
graviton KK states, with the coupling strength being $\sim 0.01 
\Lambda_\pi^{-1}$, independently of the value of $\nu$.) The precision 
electroweak bounds give constraints on gauge and graviton KK masses that 
are less than 1 TeV.  
In region II with $-0.6 <\nu <-0.5$, the fermionic couplings of the gauge KK
tower grow weaker, yielding an almost non-existent bound from precision
electroweak data.  The corresponding graviton KK tower - fermion interaction
strength increases two orders of magnitude within this range, but remains 
small. Note that constraints from the precision electroweak parameter $V$ 
disappear completely at $\nu=-0.5$, as the fermions and gauge KK states 
completely decouple at that point.
In region III, defined by $-0.5<\nu <-0.3$, the fermionic couplings of both 
the gauge and graviton towers grow rapidly and the limits from $V$ on the
masses of the first excitations lie in the few TeV range.  Lastly, in 
Region IV, corresponding to $-0.3<\nu$, the bound from $V$ is so strong 
that direct production of the KK excitations of either the gauge bosons or 
gravitons is kinematically forbidden at any planned collider. Their only 
influence in this region will be through contact interaction effects.

\nn

Before discussing the details of the collider phenomenology associated with the 
graviton and gauge KK states in these various regions, 
we note that we will assume for simplicity that 
the gauge KK states are sufficiently massive so that mixing effects can be 
neglected.  In general, the masses of the excitations of each gauge KK tower 
are given by the diagonalization of a mixing matrix, whose off-diagonal
elements are proportional to the mass of the zero-mode KK state.  Hence,
the excitations for the photon and gluon towers are automatically diagonalized 
and the masses of the KK states of the $W$ and $Z$ towers are shifted by 
$M_{W,Z}$.  This is a small effect for heavy KK states and hence
we assume that the members in the $Z$, $W$, photon and gluon towers are 
highly degenerate, level by level. This implies that the $Z$ and 
$\gamma$ tower members strongly interfere with one another appearing as a 
single resonance, $Z^{(n)}/\gamma^{(n)}$, and are hence not separable at 
colliders. This scenario is also realized in the historically more 
conventional KK gauge analyses{\cite {sminbulk,getV}} with flat spacetime. 

\begin{figure}[htbp]
\centerline{
\psfig{figure=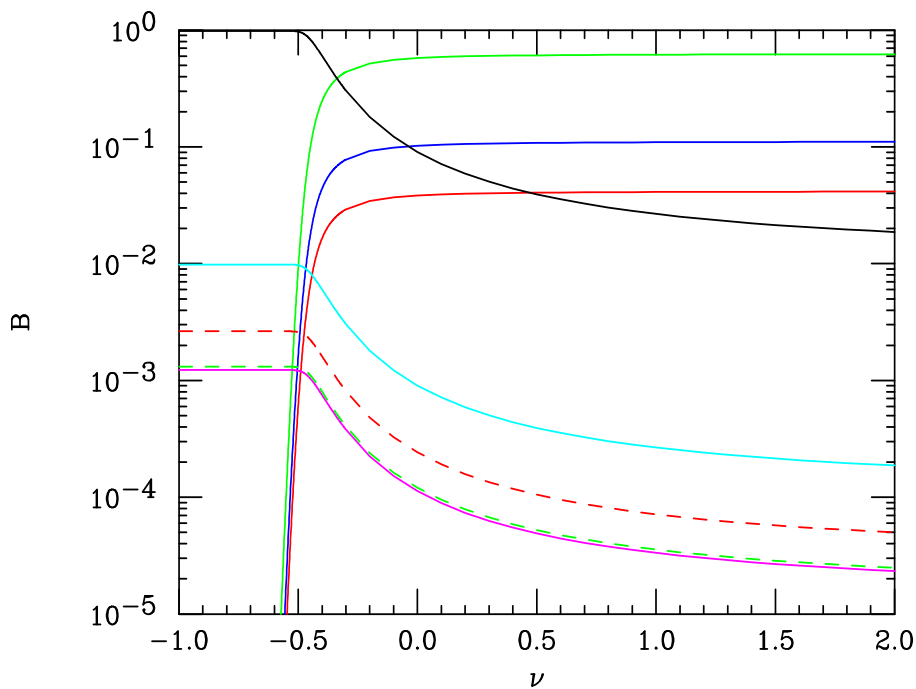,height=9cm,width=12cm,angle=90}}
\caption{Branching fractions for two-body decays of the first KK graviton 
excitation with a mass of 1 TeV as a function of $\nu$. 
The final states are, from top to bottom on the right-hand side of the figure, 
pairs of light quarks, tops, leptons, higgs, gluons, 
$W$'s, $Z$'s and photons. The Higgs mass is assumed to be 120 GeV.}
\label{fig6}
\end{figure}

It is instructive to first examine the dependence of the graviton
branching fractions on the fermion bulk mass parameter.  Figure \ref{fig6}
shows these branching fractions for the first graviton excitation with a
mass of 1 TeV.  In regions I and II, we see that the primary decay mode, by 
approximately two orders of magnitude, is that of a pair of Higgs bosons!
The decay rates into more conventional channels, such as dijets, are
uncharacteristically tiny and hence the usual signatures for graviton
production will be altered.  In regions III and IV, the fermions are no
longer decoupled allowing for large branching fractions
into fermion pairs, and thus the typical graviton production signals at
colliders become available.
We now examine the phenomenology of each region in turn.

\nn

We first consider region I. Since the fermion couplings here are far too 
weak to allow for graviton production at colliders, it is natural to ask 
whether such states could be produced via gluon-gluon fusion at the LHC since 
the $gg$ luminosity is so large at those energies. This idea runs into two 
immediate problems. First, in region I we know from the $V$ analysis and the 
mass relations in Fig. \ref{fig1} 
that the first graviton KK mass is in excess of 900 GeV. This expectation 
drastically reduces the production rate for such a heavy state 
down to the level of at most 
tens of events for a luminosity of 100 $fb^{-1}$. The 
second problem is one of signal.  As shown in Fig. \ref{fig6} 
the primary decay mode in region I is into a pair of Higgs bosons.
For more customary channels, such as dijets, we end up paying an
additional factor of 100 leaving us with no signal. We thus conclude that 
graviton KK states in region I are not observable at the LHC or any other
planned collider.

Before continuing we note that when calculating cross sections and 
production rates for the first KK graviton and gauge bosons we have assumed 
that they can decay only into SM, \ie, zero-mode states. We have found this
 to be 
a reasonable approximation for all the cases of interest to us though other 
final states may occur. One example of this possibility is
the decay of a first KK graviton excitation 
into one zero-mode gauge or fermion state together with a first 
excited mode of a gauge or fermion state. For fermions this is kinematically 
allowed only over a small range of $\nu$ but can correspondingly 
always occur for the 
asymmetric gauge final state. Such partial widths have been calculated  and
usually lead to rather small effects 
due to the reduction of the graviton coupling strength 
at the vertex and do not result in changes to the
peak cross sections by more than 
$\simeq 10-20\%$. Thus their neglect provides an adequate approximation for
the result presented here.

\nn
\begin{figure}[htbp]
\centerline{
\psfig{figure=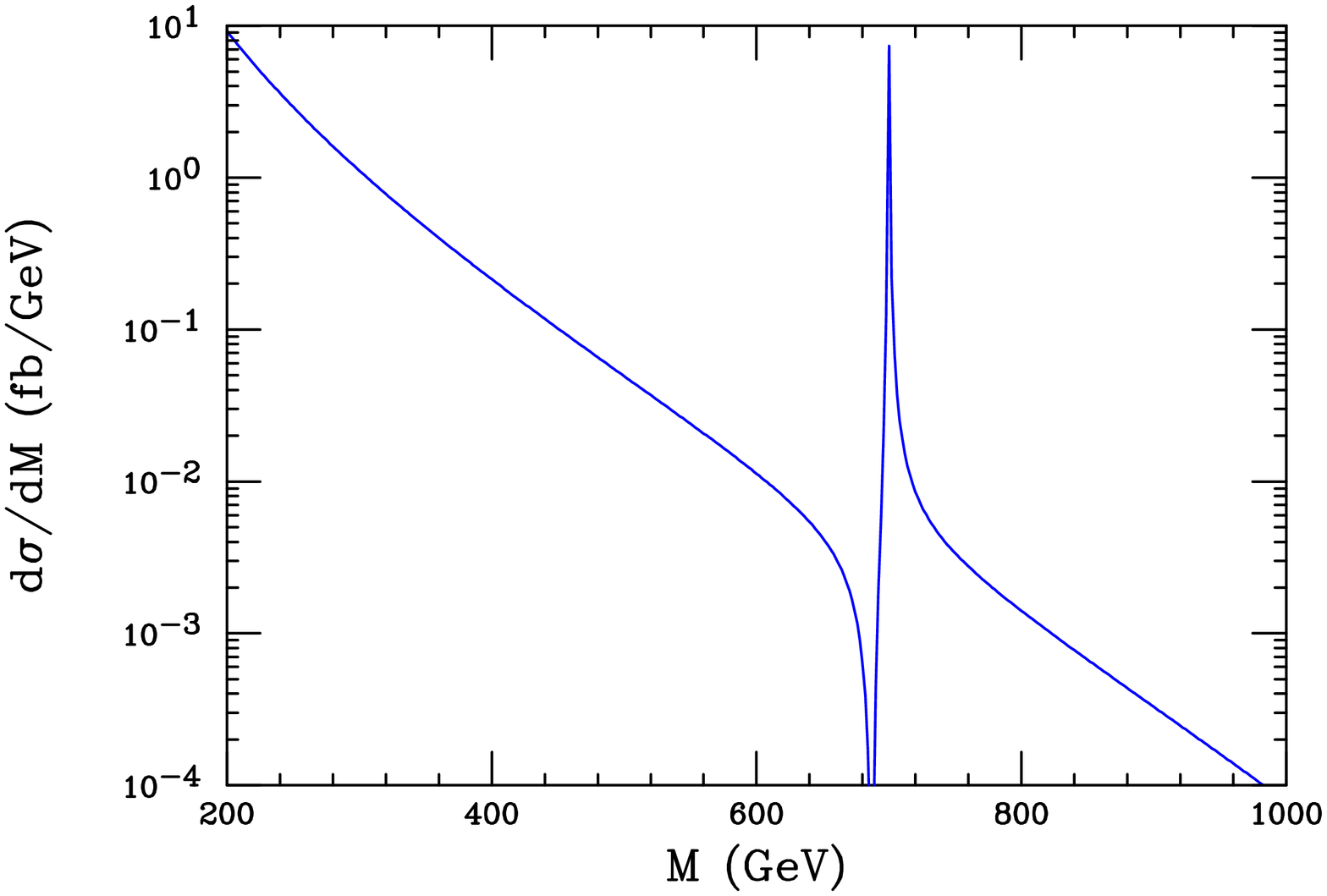,height=9.cm,width=12cm,angle=90}}
\vspace*{0.15cm}
\centerline{
\psfig{figure=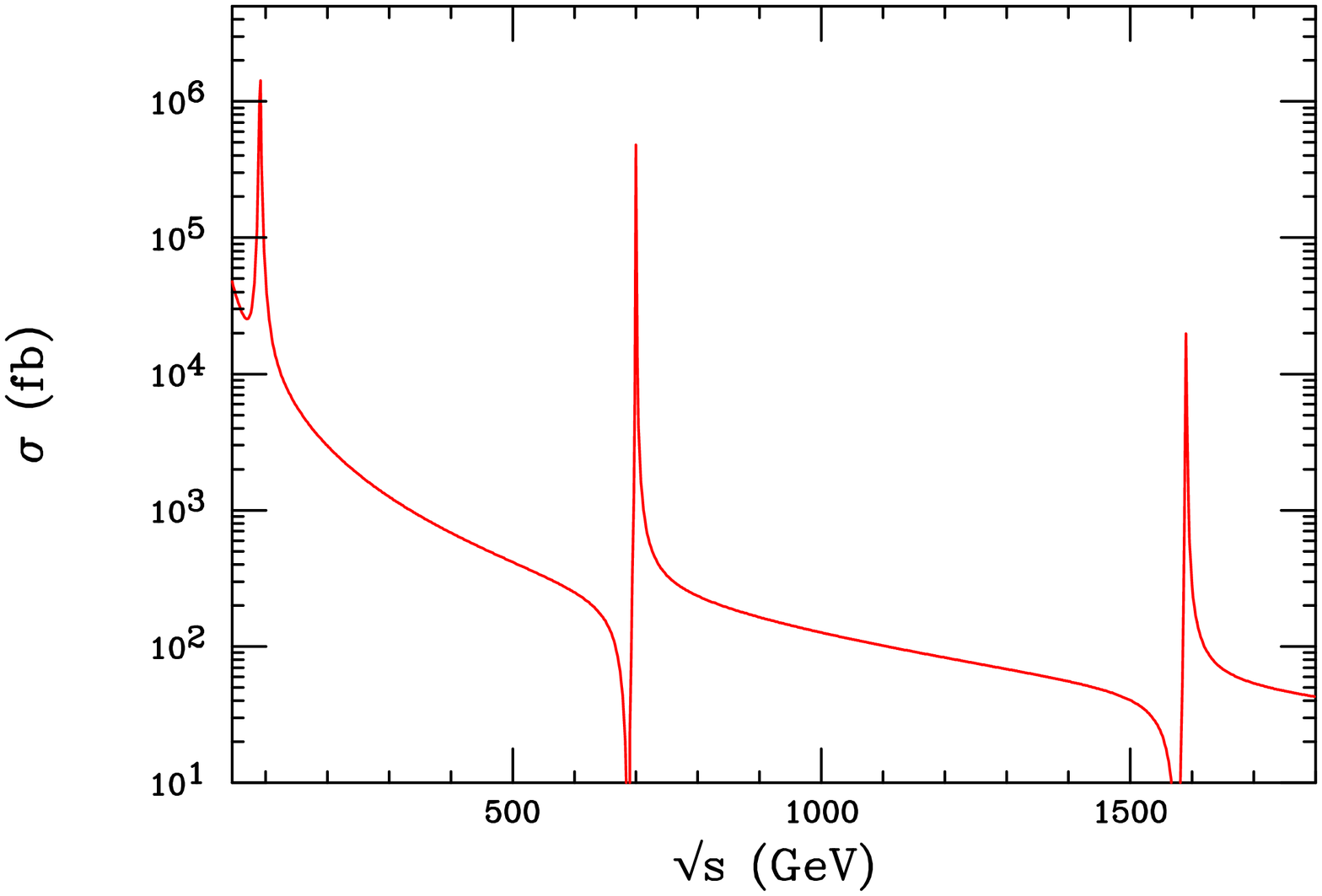,height=9.cm,width=12cm,angle=90}}
\caption{Production cross section in Region I for the first neutral KK gauge 
boson excitation with $m_1=700$ GeV in (top) Drell-Yan collisions at the 
Tevatron and in (bottom) $e^+e^-\to \mu^+\mu^-$ at a Linear Collider.  In
the latter case, the second KK gauge excitation is also displayed.}
\label{fig7}
\end{figure}

\begin{figure}[htbp]
\centerline{
\psfig{figure=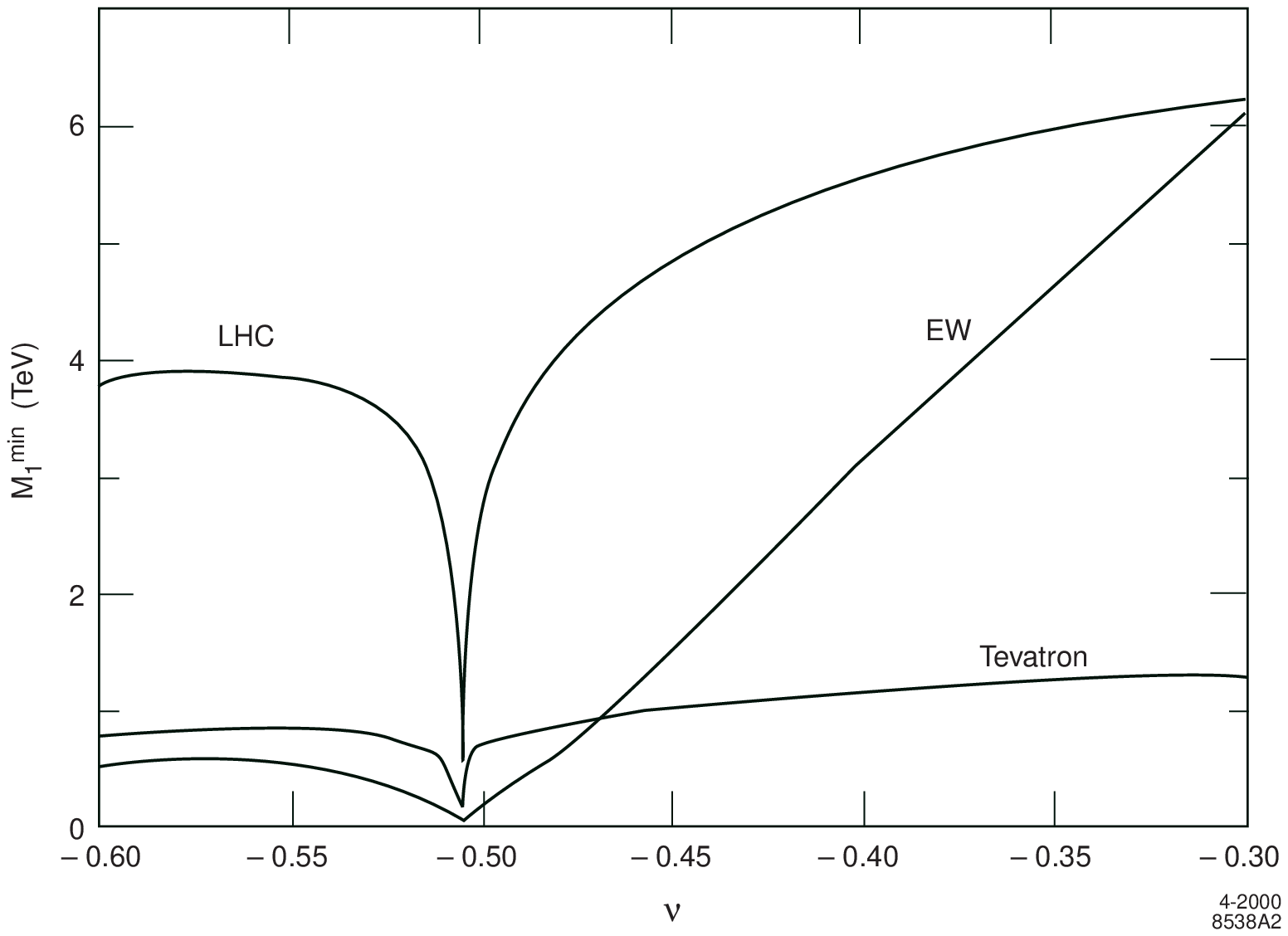,height=9cm,width=12cm,angle=0}}
\caption{Direct and indirect bounds on the mass of the first KK gauge boson in 
regions I-III. The upper (lower)most curve on the right side 
is from Drell-Yan searches at the 
LHC (Run II Tevatron) with a luminosity of $100~(2)~fb^{-1}$. The sharply 
rising curve on the right arises from the indirect radiative corrections bound.}
\label{fig8}
\end{figure}

Next, we turn to the gauge KK states; they are expected to be lighter than the 
gravitons and the lowest lying states have coupling strengths to fermions 
approximately 
$20\%$ as large as do the corresponding SM gauge bosons. However, couplings 
of this 
strength are sufficiently large as to permit significant cross sections at 
colliders as is shown in Figs. \ref{fig7}a and b for 
the Tevatron and at a Linear 
Collider, respectively. In both cases these figures show the production of a 
700 GeV $Z^{(1)}/\gamma^{(1)}$ state which has an unusually distorted 
excitation curve due to the strong interference between the $\gamma^{(1)}$ and 
$Z^{(1)}$ states and the SM $\gamma$ and $Z$ 
background exchanges. This composite excitation is 
quite narrow for its mass due to the small gauge couplings and is quite 
unlike other possible s-channel resonances such as a graviton, $Z'$ or
sneutrino. The observation of the gauge KK 
states will thus be the only signal for the RS model in this region. 
Figure \ref{fig8}
compares the search reach for these KK gauge bosons by both the Tevatron and 
LHC in the Drell-Yan channel for region I (as well as II and III) in 
comparison to the bound obtained from the $V$ analysis. Here we see that there 
is substantial room for discovering such gauge KK states with these machines 
in this region.

\nn

\begin{figure}[htbp]
\centerline{
\psfig{figure=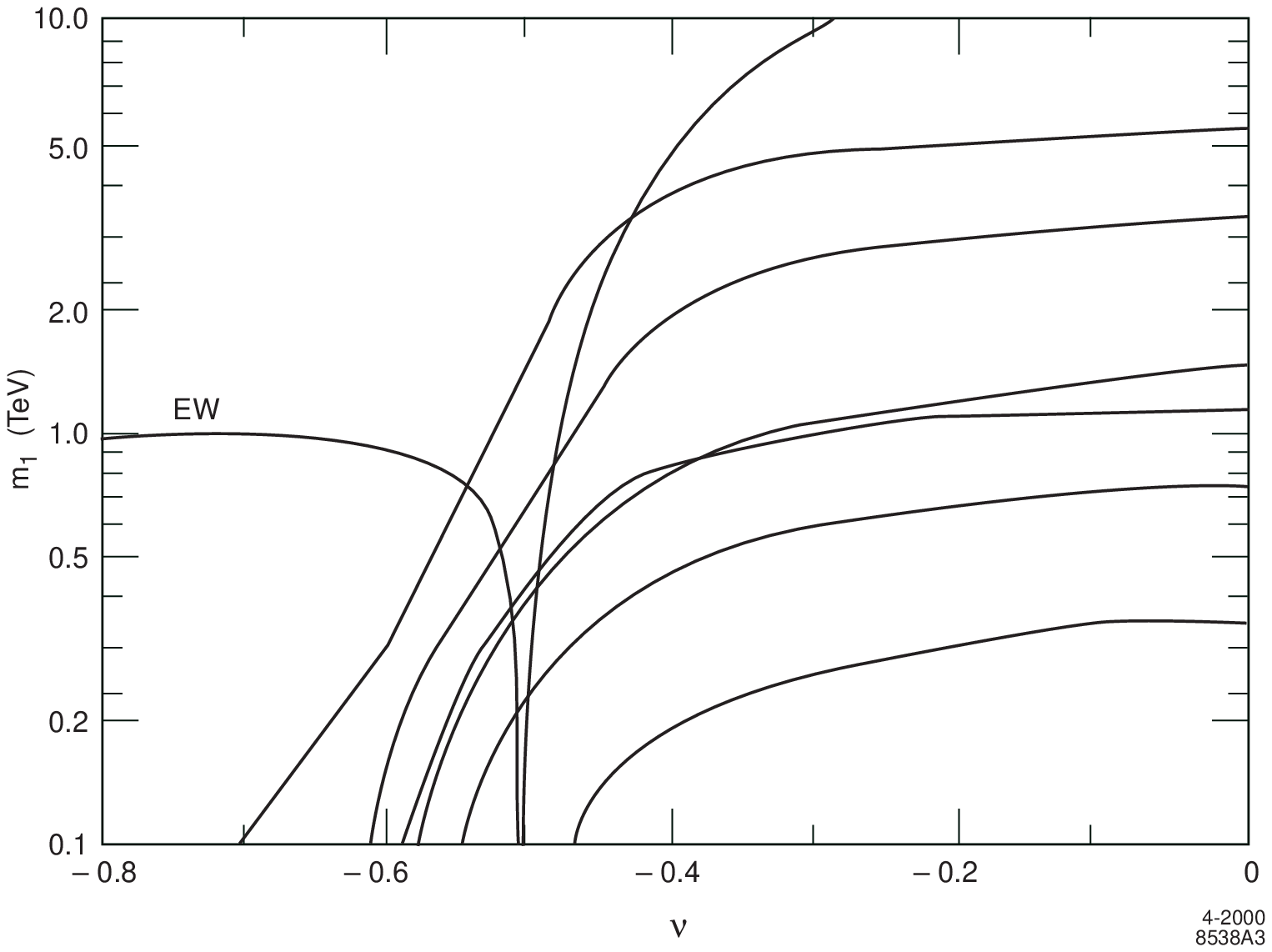,height=9cm,width=12cm,angle=0}}
\caption{Direct and indirect bounds on the mass of the first KK graviton. The 
upper (lower) set of three curves correspond to Drell-Yan searches at the LHC 
and Tevatron for the same luminosities as in the previous figure. Within each 
set of curves, from top to bottom, $\cc=1, 0.1$~and 0.01, respectively. The 
remaining curve arises from the radiative corrections bound on the gauge boson 
mass and the employs the mass relationships shown in Fig. 1.}
\label{fig9}
\end{figure}

\begin{figure}[htbp]
\centerline{
\psfig{figure=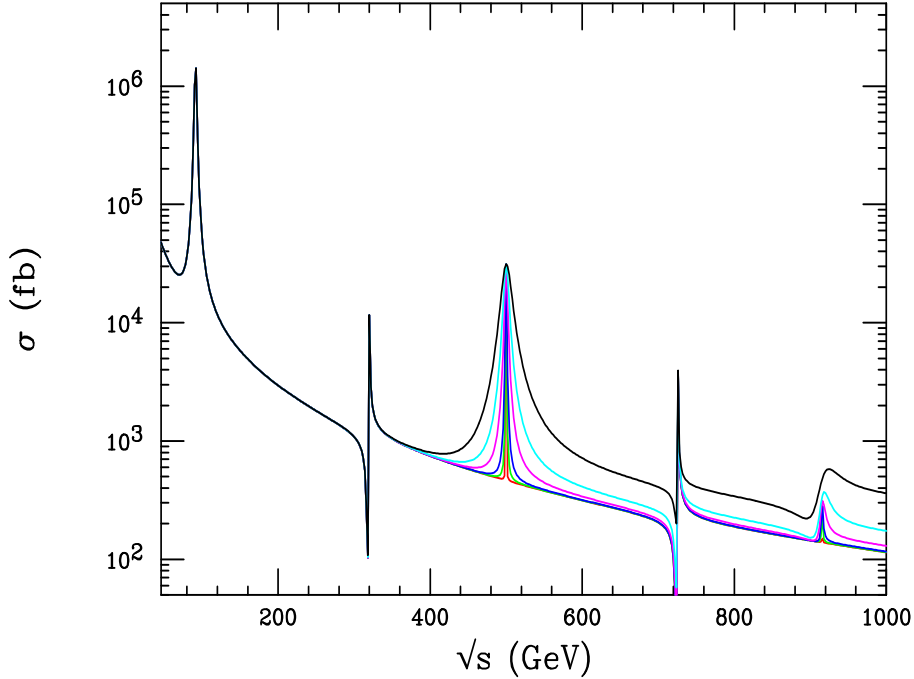,height=9cm,width=12cm,angle=90}}
\caption{Production of graviton and neutral gauge KK excitations at a linear 
collider via the process $e^+e^-\to \mu^+\mu^-$ when the fermion bulk mass
parameter is larger than -0.5 and
first graviton KK excitation is 500 GeV
for various values of \cc.}
\label{fig10}
\end{figure}

In region II with the shrinking of the gauge couplings there is a general 
degradation of the search reaches for the KK gauge bosons at both the Tevatron 
and LHC as shown in Fig. \ref{fig8}. 
Simultaneously the fermion couplings to the graviton are beginning to 
turn on and, as can be seen from Fig. \ref{fig9}, the LHC has 
some chance of producing $\sim$1 TeV gravitons for large values of 
$c=k/\mpl \geq 0.1$. Once $\nu$ exceeds $-1/2$ and we are in region III we see 
that the LHC can discover KK gauge bosons for all values of $\nu$ less than 
about $-0.3$. The window for graviton discovery, due to their larger masses is 
somewhat slimmer and is limited to larger values of $c$. When $\nu >-0.42$ 
gravitons can no longer be observed at the LHC due to their large masses. 
It is clear that in region III the KK excitations of both the graviton and 
gauge bosons can be {\it simultaneously} produced as is depicted in 
Fig. \ref{fig10} for an \epem\ linear collider.

\nn

\begin{figure}[htbp]
\centerline{
\psfig{figure=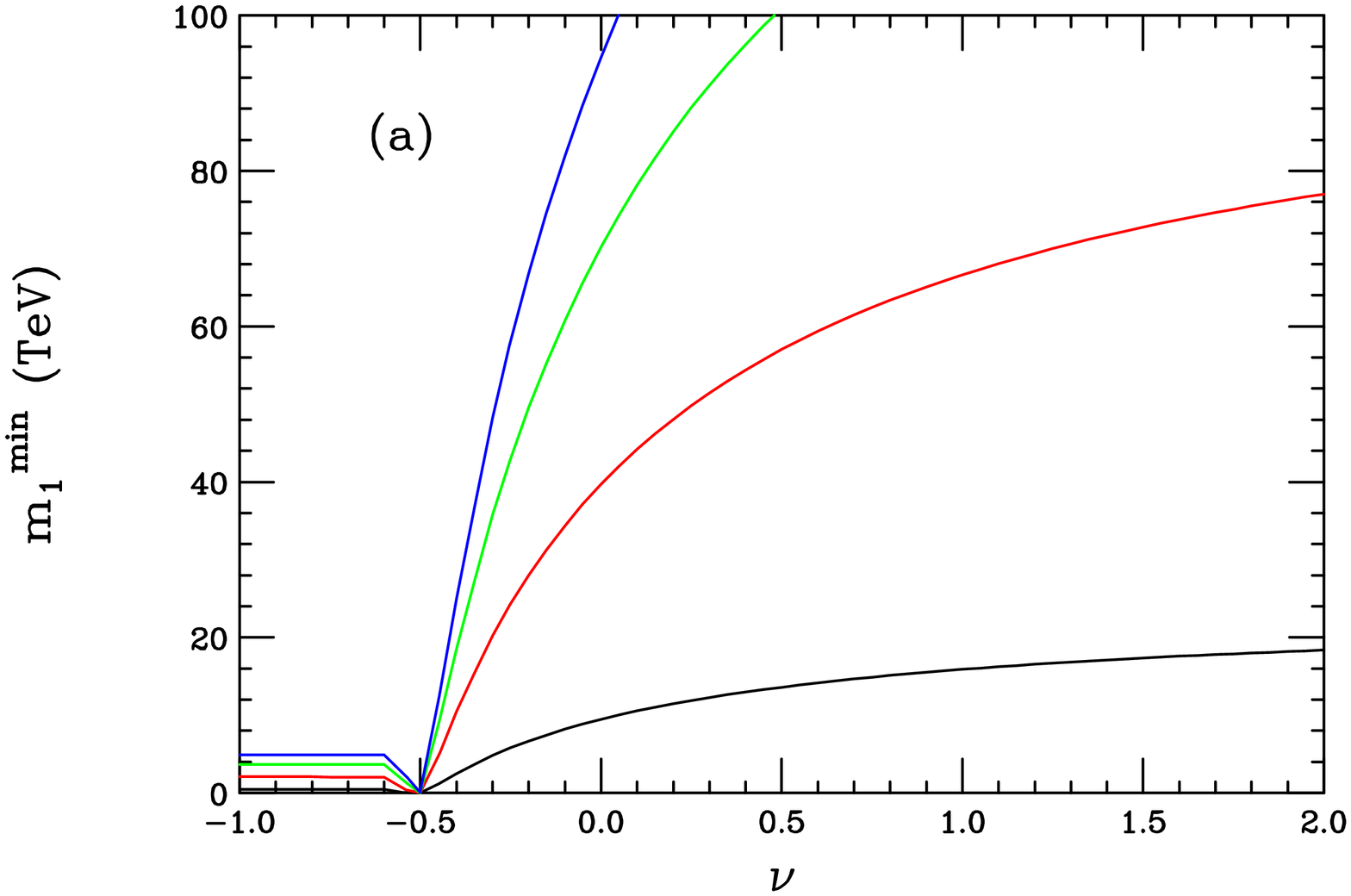,height=9.cm,width=12cm,angle=90}}
\vspace{0.1cm}
\centerline{
\psfig{figure=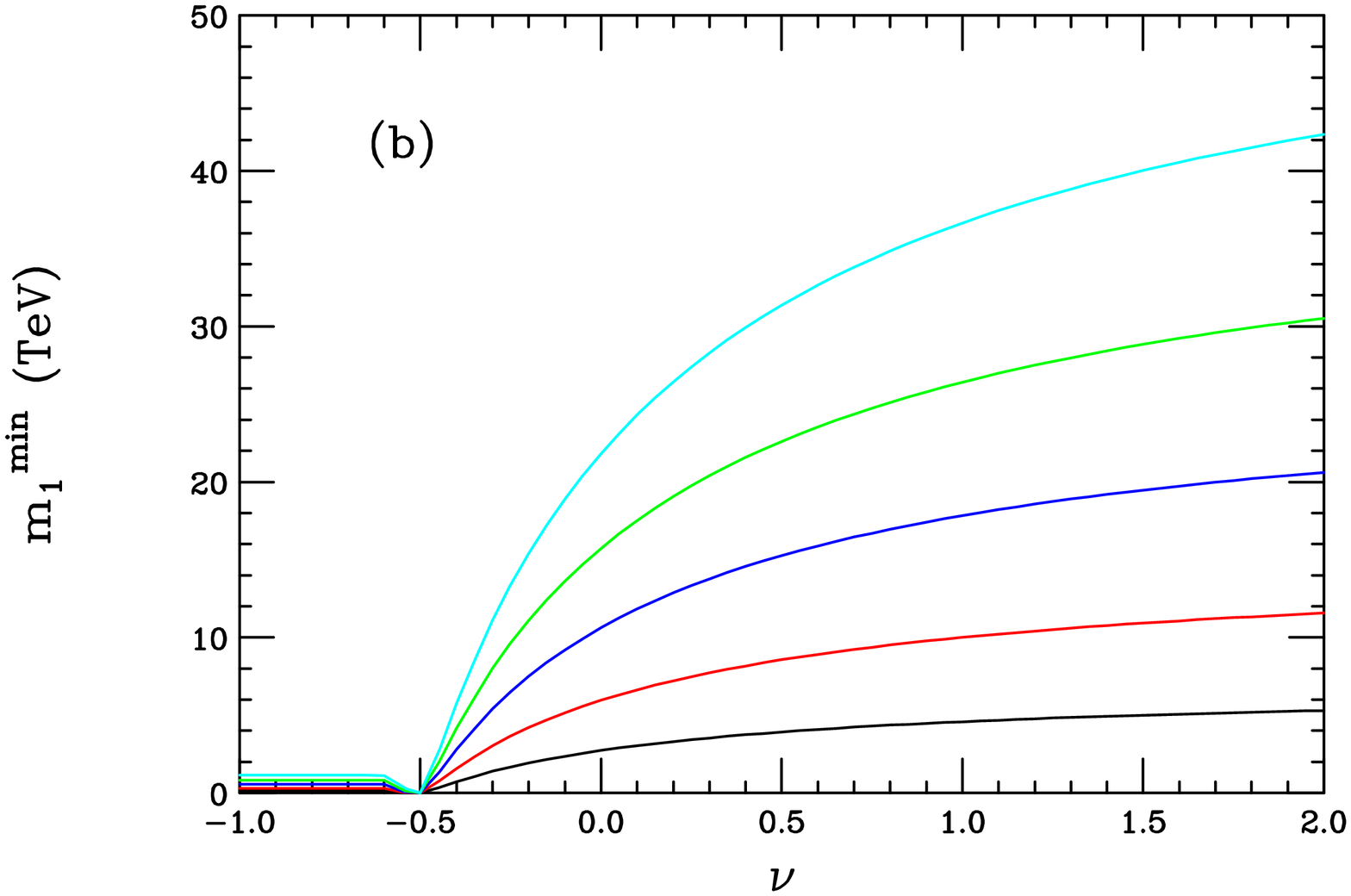,height=9.cm,width=12cm,angle=90}}
\caption{Search reach in region IV for the indirect effects of KK gauge and 
graviton exchange through contact-like interactions at (a) lepton colliders and
(b) hadron colliders.  The curves correspond from top to bottom (a) the NLC with
500 \infb\ and $\sqrt s=$ 1500, 1000, and 500 TeV, and LEP II at $\sqrt s=
195$ GeV with 1 \infb; (b) the LHC with 100 and 10 \infb, and the
Tevatron at Run II with 30 and 2 \infb, and the Tevatron at Run I.}
\label{fig11}
\end{figure}

In region IV the precision electroweak constraints show that the first 
excitation of both the gauge and graviton KK towers is above the kinematic
threshold for direct production at the LHC.  However, their contribution
to fermion pair production may still be felt via virtual exchange, similarly
to contact-like interactions.  These effects are dominated by the gauge KK
tower exchange as the gauge KK states are lighter, level by level, and much
more strongly coupled than the corresponding KK gravitons.  In addition,
the gauge KK tower contributes to fermion pair production via a dimension-six
operator, whereas the graviton contribution is dimension-eight.  The effects
of the KK graviton exchange can thus be essentially neglected in comparison
to the KK gauge contributions.  We modify the results of 
Refs. \cite{getV,tev,had} to include the effects of KK tower exchange and
present the resulting 95\% C.L. search reach in Fig. \ref{fig11} for various
lepton and hadron colliders with center-of-mass energies and integrated
luminosities as indicated.  All fermion final states were employed in the
lepton collider analyses, while only Drell-Yan data was included in the hadron
case.  We see that the LHC with 100 \infb\ will give
comparable bounds to those obtained from our precision electroweak analysis,
while the NLC has a substantial search reach.
These bounds, as well as those shown in Fig. \ref{fig4}, demonstrate that this 
is a problem region for the RS model as they naturally lead to values of 
$\Lambda_\pi$ significantly in excess of 10 TeV.

\nn

\section{Phenomenology of Wall Fields}

From the discussion in the previous section it is clear that if the SM fields 
propagate in the RS 
bulk then there is only a small range of $\nu$ for which the 
RS model can be directly tested through the production of graviton 
resonances. Either such states are constrained to be too massive to be 
produced, as can be inferred from the analysis of precision electroweak data,
or they decouple from the zero-mode fermions and cannot 
be produced at all. In addition, the value of $\Lambda_\pi$ is allowed to 
be $\lsim 10$ TeV only in regions I-III, corresponding to
the range $-0.9$ to $-0.8\lsim\nu\lsim -0.3$.
For larger values of the fermion bulk mass parameter, which is most of 
this parameter's  
natural range, the lower bounds on $\Lambda_\pi$ begin to approach 100 
TeV.  One may argue that this is disfavored since it is so far away from the 
weak scale and may create additional hierarchies.
Thus unless one can construct a model wherein the value of $\nu$ 
naturally lies in the above narrow range it appears that placing
the SM in the RS bulk is somewhat undesirable.
For this reason, and to complete our earlier brief 
analysis{\cite {dhr}}, we now explore the phenomenology for the case where 
the SM field content is entirely confined to the TeV-brane. 

We remind the reader that in the case where only gravity propagates in
the bulk, the graviton KK tower couplings to all
wall fields, and for all tower members $n\geq 1$, 
are simply suppressed by \lpi; the zero-mode coupling remains Planck
scale suppressed.  In the language developed in section 2, this corresponds
to values of the coefficients, $C^{f\bar fG}=C^{AAG}=1/e^{-kr_c\pi}$.

\subsection{Bounds from the Oblique Parameters $S$, $T$, and $U$}

In addition to both direct and indirect searches for new physics at
colliders, precision measurements can also provide useful constraints on
new interactions\cite{htt}.  We saw above that a detailed analysis of
radiative correction effects parameterized by the quantity $V$ gave 
powerful bounds on the mass of the first graviton excitation when the SM
gauge fields (and fermions) were in the bulk.  However, in the case where
the SM completely resides on the 3-brane, it is clear that the masses of the 
bulk graviton
fields are no longer correlated to $V$ at tree-level, so that this analysis is
no longer useful in obtaining constraints.

A different approach to probing deviations in electroweak data due to new
physics is through shifts in the values of the oblique parameters $S$, $T$, and
$U$\cite{tatsu}.  In the case of graviton KK towers, it is clear that loops
involving such particles will contribute to the transverse parts of the SM
gauge boson self-energies, which will then reveal themselves in deviations
in $S$, $T$, and $U$.  Recently Han, Marfatia, and Zhang\cite{han} have
considered the graviton tower contribution to these parameters within the
context of the ADD scenario arising from both seagull and rainbow diagrams.  
This analysis can be modified in a relatively
straightforward fashion to the case of localized gravity by recalling
(i) that the coupling strength of the graviton tower is inversely proportional
to $\Lambda_\pi$ and not \mpl, and (ii) the masses of the RS KK states
are widely separated so that the sum over them must be performed explicitly
and cannot be performed via integration.  Since gravity becomes strong
for momenta greater than the scale \lpi, we must introduce an explicit
cut-off, $M_c=\lambda\lpi$ with $\lambda\sim{\cal O}(1)$, to render 
the integrals and sums finite.  For practical purposes we perform all of
the integrations analytically leaving only the KK tower sum to be
performed numerically by making use of the relations $\lpi=m_1^{\rm grav}\mpl/k
x_1^{G}$ and $m_n^{\rm grav}=m_1^{\rm grav}x_n^{G}/x_1^{G}$.  For example,
the seagull diagram yields the simple result
\be
\Pi(p^2)={\lambda^2p^2\over 48\pi^2}\sum_n y_n^{-2}\left[ {1\over 3}
+4y_n+10y_n^2+10y_n^3\ln{y_n\over 1+y_n}\right]\,,
\ee
where $y_n\equiv(m_n^{\rm grav}/M_c)^2$.  Unlike the ADD case, the resulting
values for the shifts in the oblique parameters are found to be
only proportional to $\lambda^2$ instead of $\lambda^4$; 
we set $\lambda=1$ in our numerical results
below.

\begin{figure}[htbp]
\centerline{
\psfig{figure=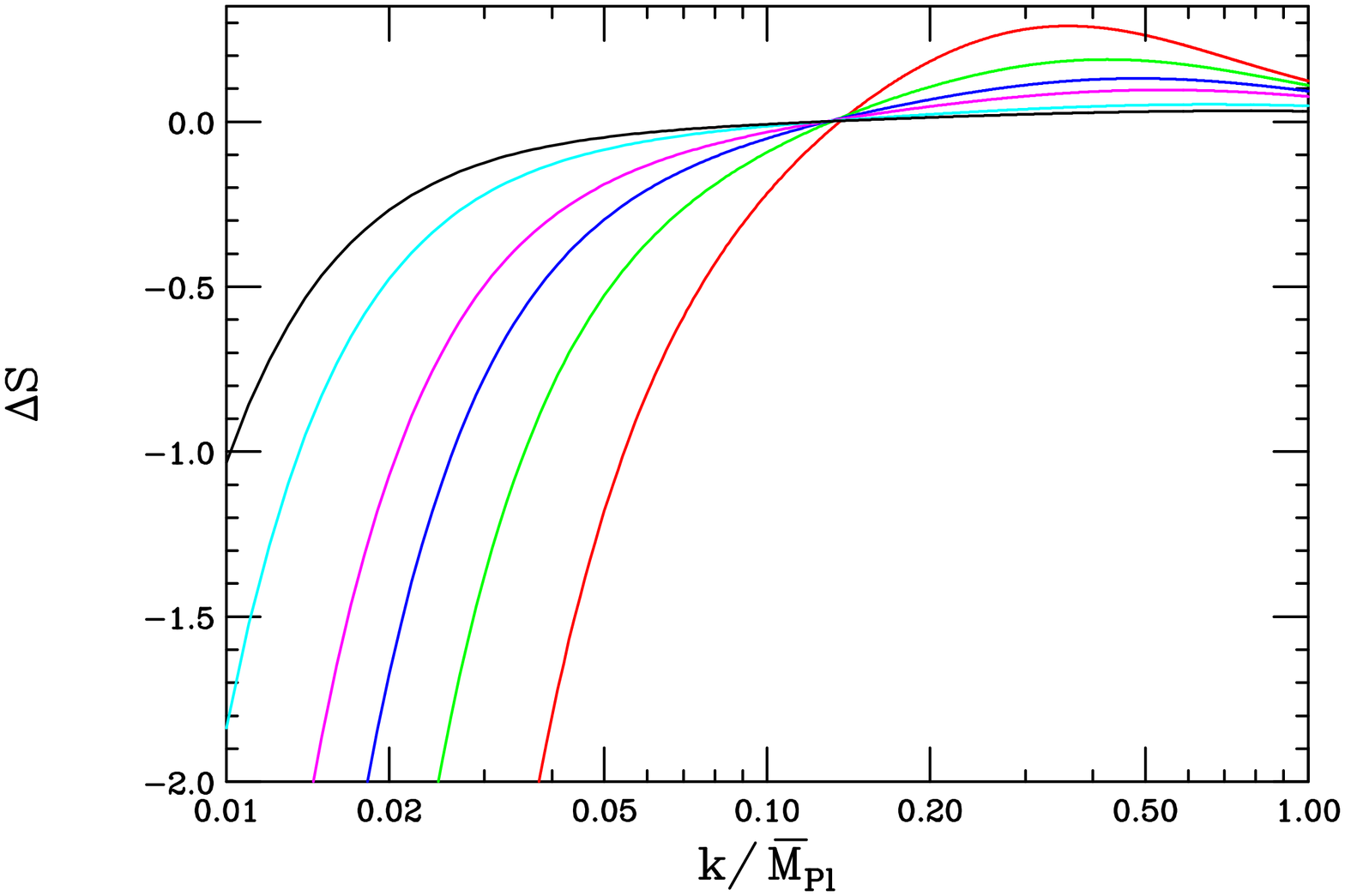,height=9.cm,width=12cm,angle=90}}
\vspace*{0.25cm}
\centerline{
\psfig{figure=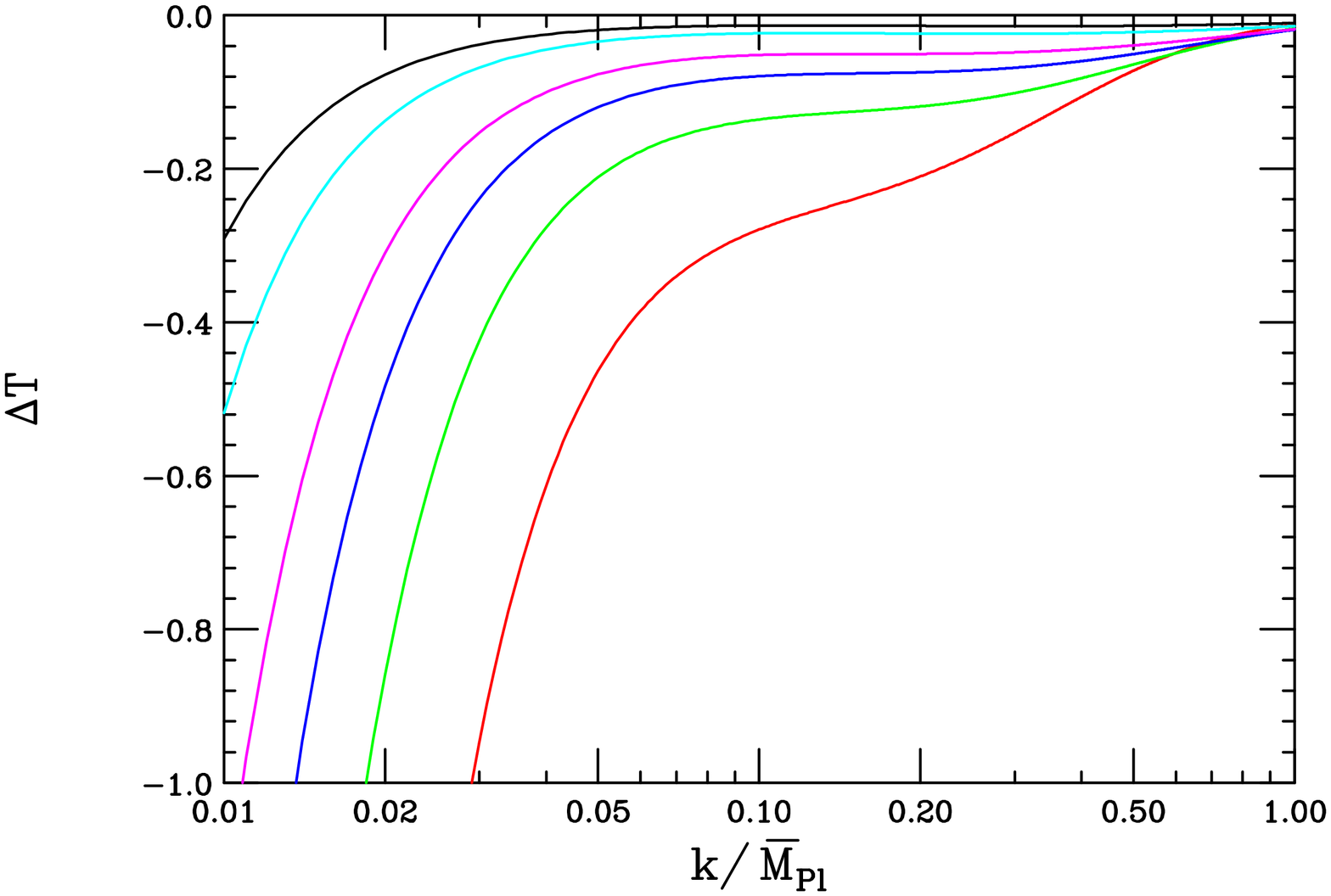,height=8.cm,width=8cm,angle=90}
\psfig{figure=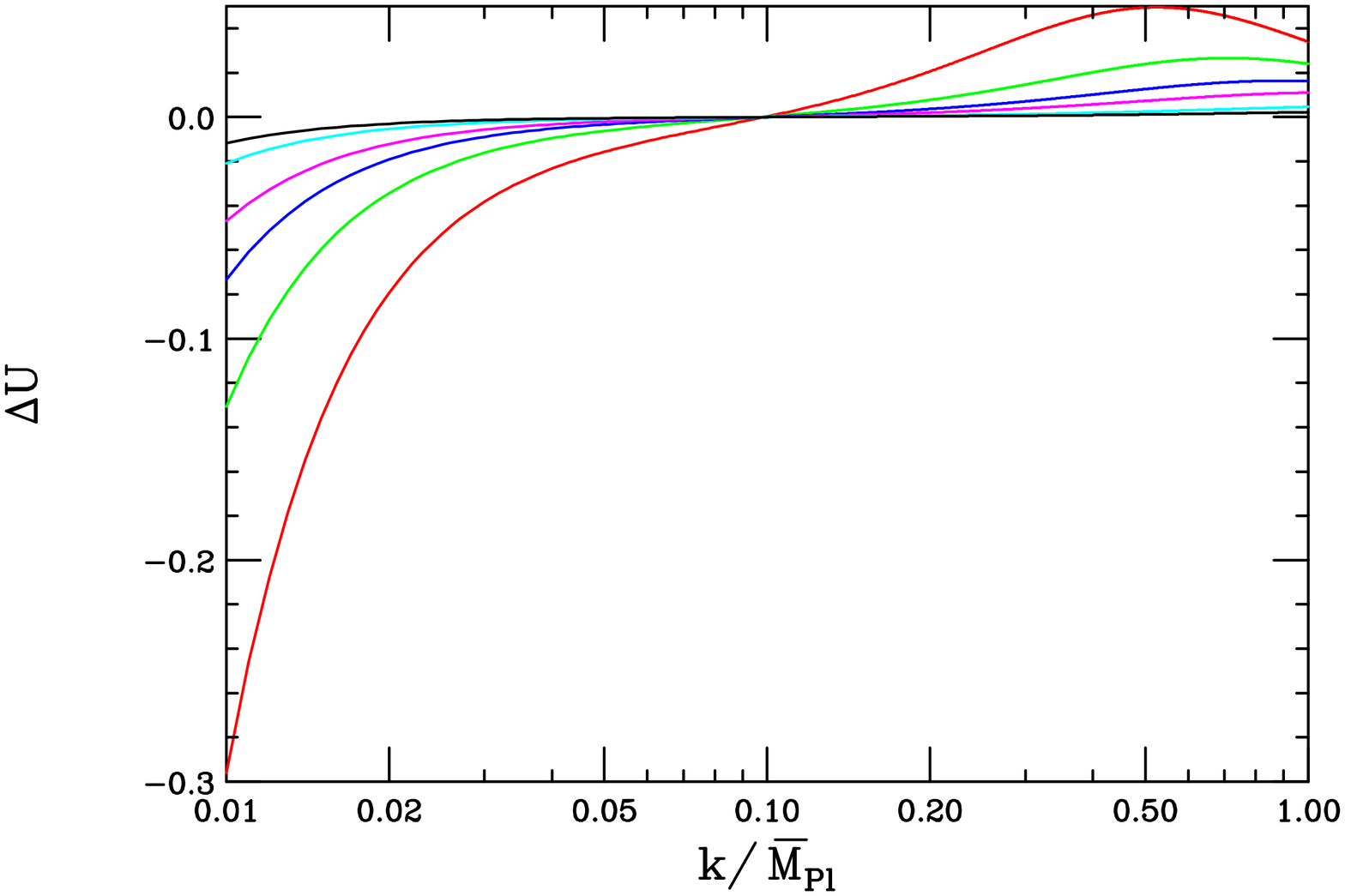,height=8.cm,width=8cm,angle=90}}
\vspace*{0.25cm}
\caption{Shifts in the oblique parameters $S$, $T$, and $U$ 
as functions of \cc\ when the 
SM resides on the TeV-brane.  From bottom to top the curves correspond to 
$m_1^{\rm grav}=200,\, 300,\, 400,\, 500,\, 750$, and 1000 GeV.}
\label{fig12}
\end{figure}

\begin{figure}[htbp]
\centerline{
\psfig{figure=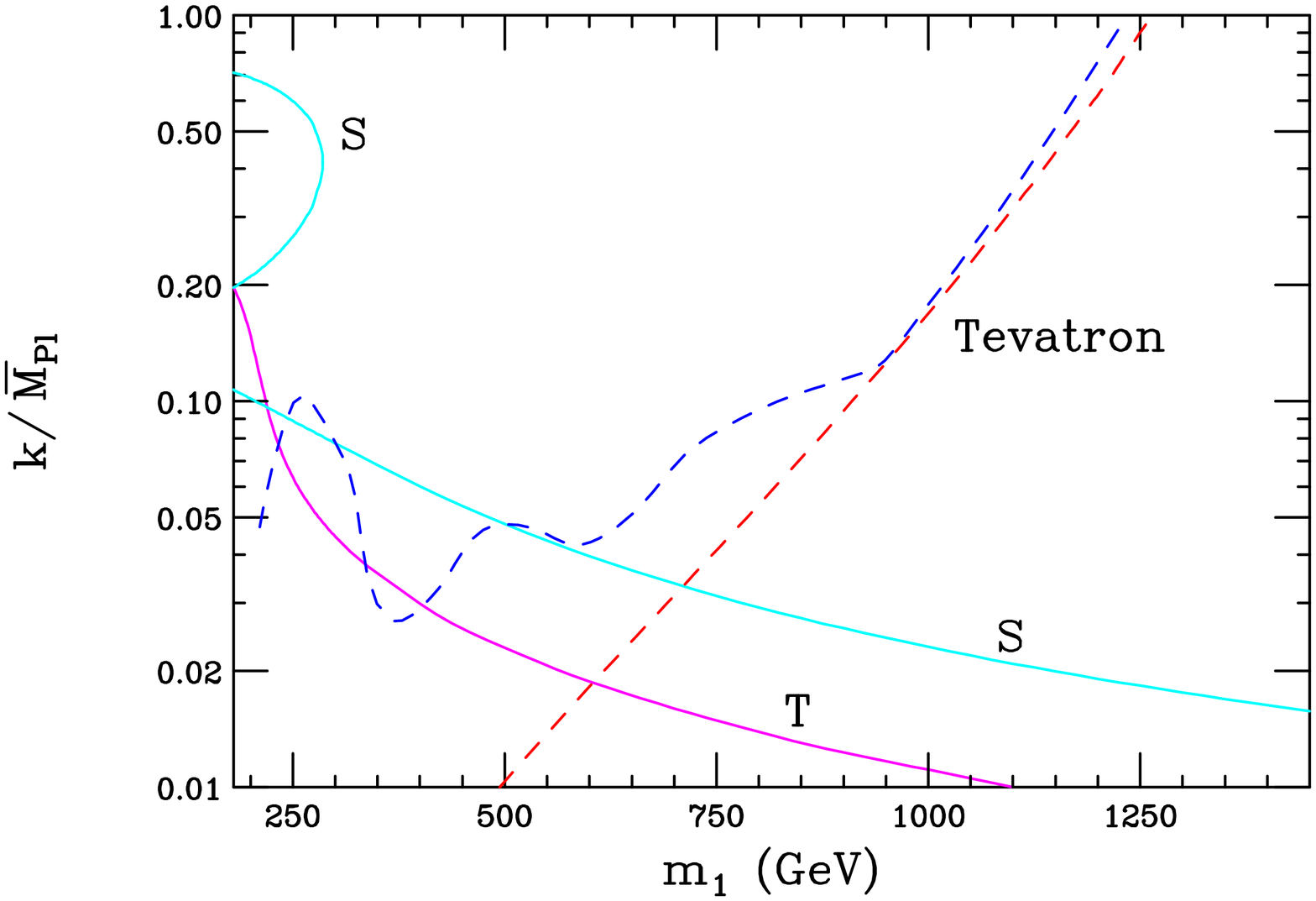,height=9cm,width=12cm,angle=90}}
\vspace*{0.25cm}
\caption{Excluded regions in the $\cc-m_1^{\rm grav}$ 
plane for gravitons coupling to SM 
fields on the wall. The purple and light blue curves arise from oblique 
corrections constraints ($T$ and $S$, as labeled) and excluded regions are 
below and to the left of these curves. The dark blue bumpy dashed 
and red straight dashed curves are bounds 
from Run II ($2~fb^{-1}$) Tevatron from dijet and Drell-Yan searches, 
respectively and will exclude regions above them and to the left.}
\label{fig13}
\end{figure}

Figures \ref{fig12}(a-c) display the shifts in the oblique parameters as a
function of $k/\mpl$ for various values of $m_1^{\rm grav}$.  Using the latest
values of $S$ and $T$ from a global fit to the electroweak data\cite{morris}
given by
\bea
S & = & -0.04\pm 0.10 \,,\nonumber\\
T & = & -0.06\pm 0.11 \,, 
\eea
we obtain the 95\% CL constraints in the $\cc - m_1^{\rm grav}$ plane shown in
Fig. \ref{fig13}.  Most of the excluded region arises from too large of
a negative contribution to either $S$ or $T$ from graviton loops, while the
small nose-like region along the vertical axis is eliminated by values of
$S$ which are positive and 
too large.  Note that, as usual, the parameter $U$ does not provide
a meaningful bound since it is quite small in magnitude in comparison to
$S$ and $T$.  As we can see from the figure, these constraints complement those
from direct collider searches, \eg, those at the Run II Tevatron. In fact, 
by combining the two sets of constraints we would find that a major part of 
the displayed parameter space would be excluded if nothing was found by the 
Tevatron during Run II. (Of course, the true size of the model parameter space 
is larger than what is shown in this figure.) This region would be further 
reduced in area by about a factor of two if we also required both that 
$\Lambda_\pi <10$ TeV and that the magnitude 
of the bulk curvature be less than the 5-d Planck scale as discussed in our 
earlier work{\cite {dhr2}}, which demands that $\cc$ 
be less than $\approx 0.1$. 
As will be discussed below, combining all of these requirements one 
can in fact show that the allowed 
region actually {\it closes} at graviton masses in the range near 4 TeV. 
This shows the strong interplay between data from 
precision measurements, direct collider searches, and our theoretical 
prejudices.

\nn

\nn

\subsection{Collider Phenomenology}

We now examine the direct production of the graviton KK states at high
energy colliders in the scenario where the SM fields are constrained to the
TeV-brane.  We expand on our previous work\cite{dhr} by investigating the
possibility of reasonably light graviton excitations, \eg, 
$m_1^{\rm grav}\lsim 200$
GeV.  These may have previously escaped detection at the Tevatron by having
an extremely narrow width.  In addition it is possible that their 
contributions to the oblique parameters discussed above may be cancelled by
the effects of other sources of new physics and hence this window should 
also be probed by direct collider searches.  We then turn to the more likely
scenario where the mass of the first graviton excitation is at least a few 
hundred GeV, and explore its resonance production at future colliders in 
detail.

\begin{figure}[htbp]
\centerline{
\psfig{figure=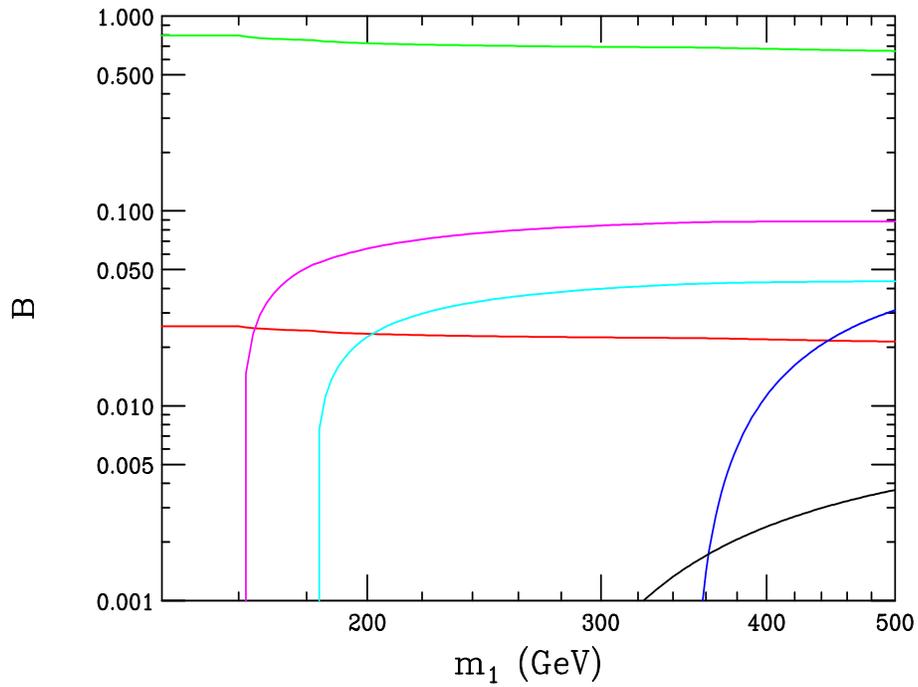,height=9cm,width=12cm,angle=90}}
\caption{Mass dependencies of the two-body branching fractions for the first 
graviton KK state in the case where the SM fields are on the wall.
From top to bottom on the right side 
of the figure the curves are for dijets, $W$'s, $Z$'s, tops, dileptons and 
Higgs pairs assuming a Higgs mass of 120 GeV.}
\label{fig14}
\end{figure}

To fully explore this phenomenology, we first determine the branching 
fractions for the decay of the first graviton KK state into two-body channels.
These are displayed in Fig. \ref{fig14} as a function of the graviton mass.  
We see from the figure that dijet final states, \ie, light quark and gluon 
pairs, dominate the graviton decays.  The leptonic channel, which yields
the cleanest signature, has a branching fraction of order a few percent
for all values of $m_1^{\rm grav}$.  Note that the branching fractions are
independent of the parameter \cc, as expected.

\subsubsection{Production of Light Gravitons}

In our earlier consideration of graviton tower phenomenology we concentrated
on the case where the first tower member was more massive than 
about $\simeq 200$ GeV. The reasons for this were two-fold: first, such masses 
are outside the range directly accessible to LEPII and, second, the Tevatron 
collider bounds for new resonances in either the Drell-Yan or dijet channel 
are essentially absent below $\simeq 200$ GeV.

\begin{figure}[htbp]
\centerline{
\psfig{figure=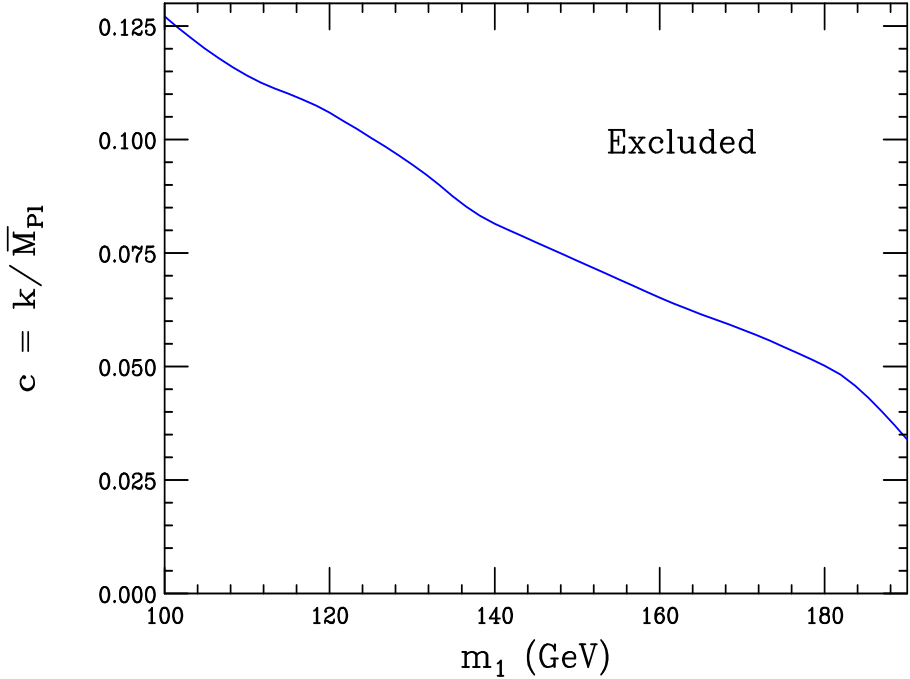,height=9cm,width=12cm,angle=90}}
\caption{$95\%$ CL upper bound on $c$ as a function of the first KK graviton 
mass from the $\tilde \nu$ bound discussed in the text.  The allowed region
lies below the curve.}
\label{fig15}
\end{figure}

There are two ways to probe this mass range below 200 GeV. The first 
possibility is to search for a narrow $s-$channel resonance in the LEPII data 
above the $Z$-pole in, for example, $e^+e^-\to \mu^+\mu^-$. Such an analysis 
has indeed been performed by the OPAL Collaboration{\cite {opal}} in their
search for $R$-parity violating $\tilde \nu_\tau$ production. The result of 
their null search is a constraint on the $R-$parity violating Yukawa coupling, 
$\lambda$, as a function of the $\tilde \nu_\tau$ mass. Clearly, this  
search can be modified to probe for narrow gravitons and a straightforward 
translation is possible; we find that
\be
c_{\rm bound}=\lambda_{\rm bound}\left[ B_\ell^{grav}x_1^2\right]^{-1/2}\,,
\ee
where $c=k/\mpl$, $x_1$ is the smallest non-zero root of the Bessel function
$J_1$ and 
$B_\ell^{grav}$ is the leptonic branching fraction of the first graviton KK 
state. The result of this analysis can be seen in Fig. \ref{fig15} where we 
observe that the bound on $c$ as a function of the first KK graviton mass is 
unfortunately rather weak. We expect, however, that these bounds should 
improve significantly by the end of the LEPII run.  Note that this direct
search supplements the constraints obtained from the oblique parameter
analysis discussed above.

\nn

\begin{figure}[htbp]
\centerline{
\psfig{figure=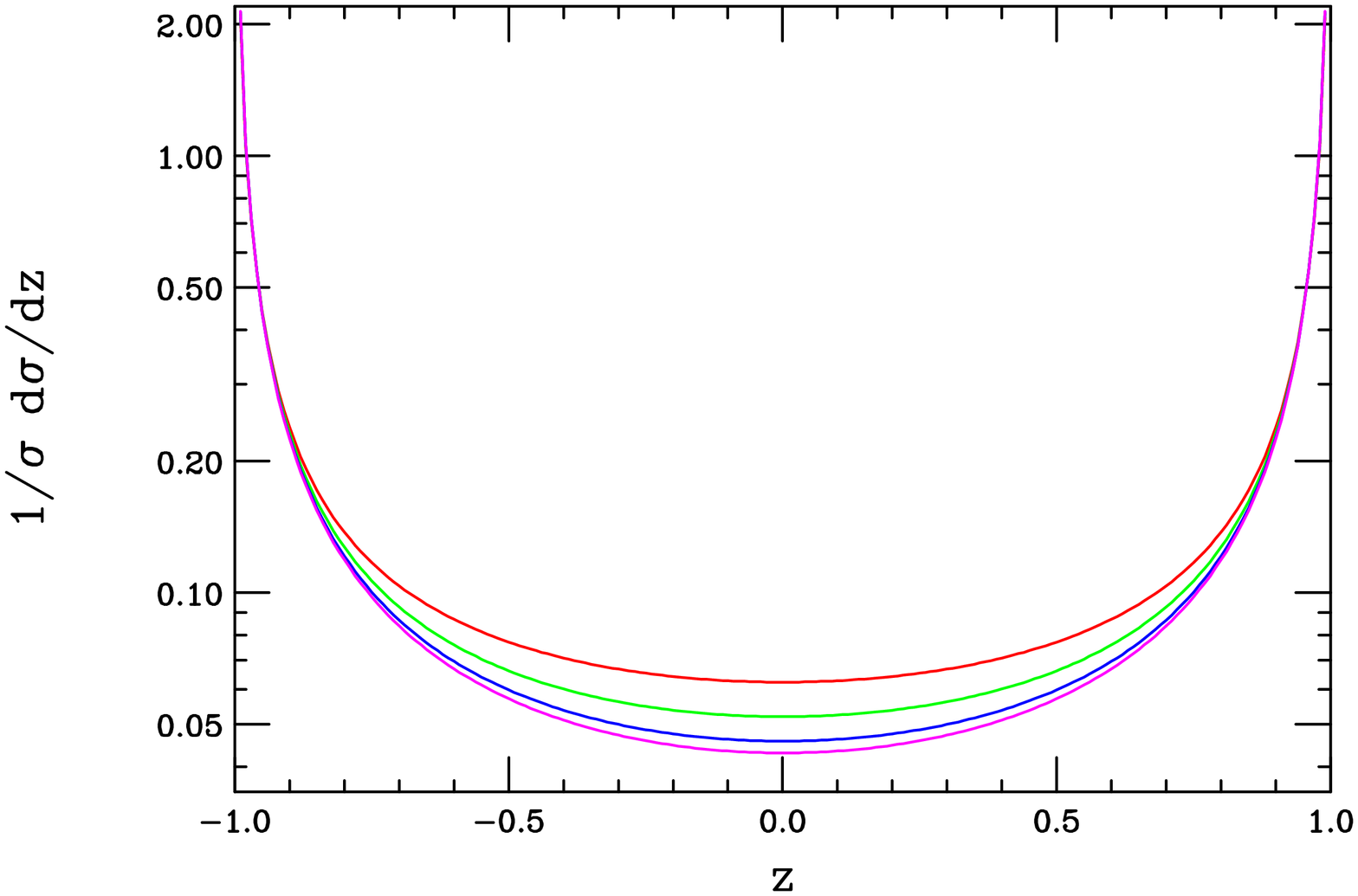,height=9.cm,width=12cm,angle=90}}
\vspace*{0.15cm}
\centerline{
\psfig{figure=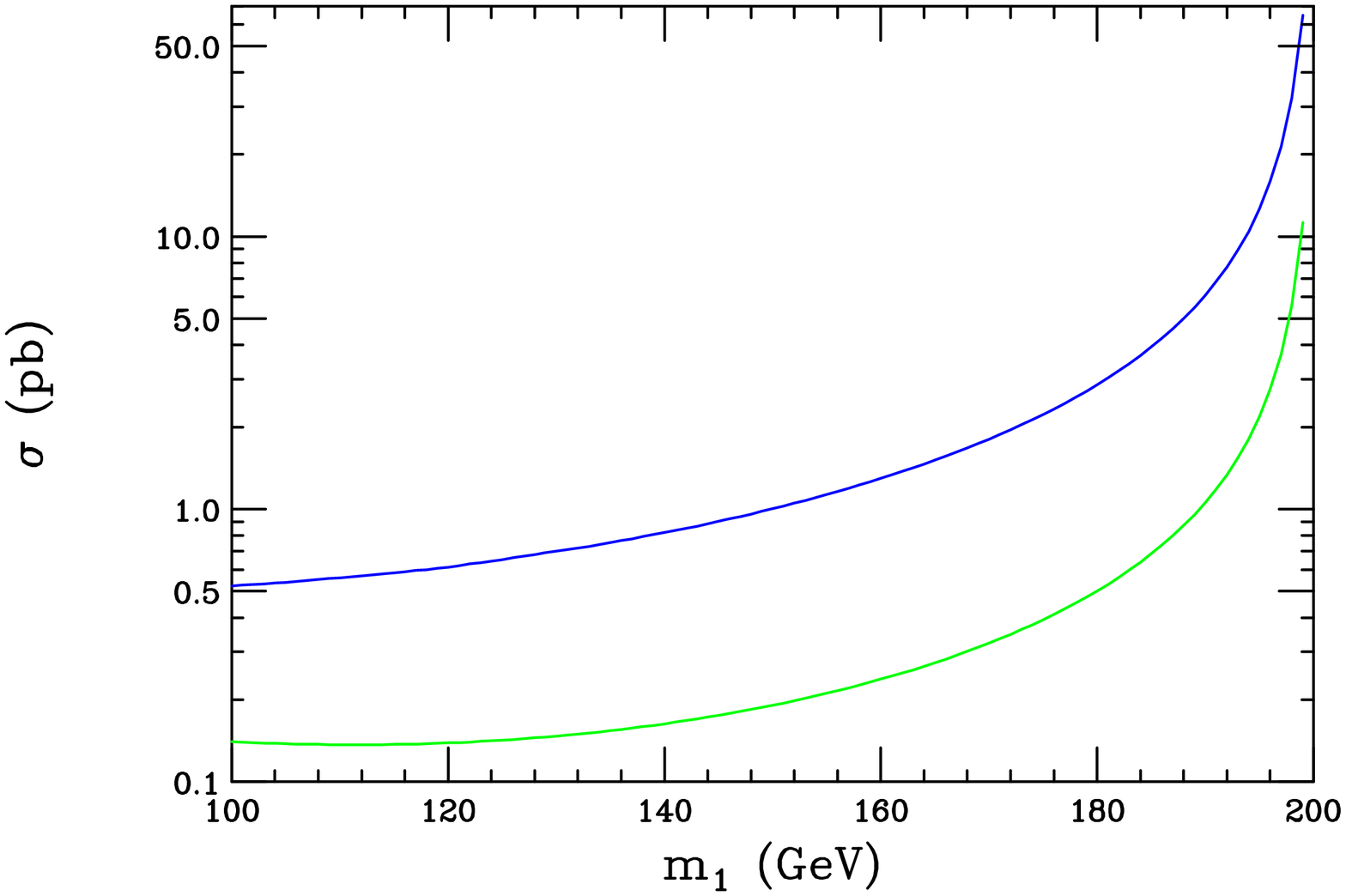,height=9.cm,width=12cm,angle=90}}
\caption{Angular distribution (top) and total cross section (bottom) for the 
process $e^+e^-\to\gamma+ G^{(1)}$ assuming $\sqrt s$=200 GeV and $\cc=0.01$. 
In the top panel, from top to bottom the curves are for a graviton
mass of 130, 150, 170 and 190 GeV, 
respectively. The lower curve in the bottom panel is the result after 
employing a cut of $15^o$ between the photon and initial electron direction.}
\label{fig16}
\end{figure}

A second possibility is to search for light gravitons by associated production 
with a photon, \eg, $e^+e^-\to \gamma+ G^{(1)}$. 
In the ADD model{\cite {nima}}, 
a number of authors have considered using this process to constrain the 
higher dimensional Planck scale as a function of the number of extra 
dimensions through a somewhat similar search process{\cite {pheno}}. In the 
ADD case, however, a tower sum of KK gravitons up to kinematic limit is also 
required so that the final state no longer appears to be resulting from an 
underlying two-body process. Unlike the ADD case,  in the RS model 
this process is a true two-body reaction leading to a mono-energetic photon 
with a differential cross section given by\cite{pheno}
\be
{d\sigma\over dz}={\alpha c^2x_1^2\over 16(1-x)}\left[ (1+z^2)(1+x^4)
+(1-3z^2+4z^4){1+x^2\over 1-z^2}+6x^2z^2\right]\,,
\ee
where $x=m_1^2/s$, $z=\cos \theta$ and $m_1$ is the mass of the lightest KK 
graviton. The production signature for this process is the mono-energetic 
photon and the decay products of the on-shell massive graviton, \eg, a pair 
of dijets, $\ell^+\ell^-$ or another $\gamma \gamma$ pair that reconstruct to 
the mass of the graviton. Given the expression above one might imagine that 
the differential distribution of photons is highly peaked in both the forwards 
and backwards directions independent of the value of $m_1$ above the $Z$ 
mass. Fig. \ref{fig16}a 
explicitly shows the resulting normalized angular distribution 
of the photon for $\sqrt s=200$ GeV and several distinct values of $m_1$ with 
the anticipated strong forward-backward peaking. Unfortunately, the continuum 
SM background from single-photon radiation has a very similar angular 
distribution but is not mono-energetic. In either case the signal to 
continuum background ratio can be somewhat enhanced by imposing a hard cut on 
the photon production angle relative to the incident electron beam. 
Fig. 16b 
shows the total integrated cross section for the process of interest as a 
function of $m_1$ both with and without the photon angular cut, assuming that 
$c=0.01$ and $\sqrt s=200$ GeV. Here we see that reasonable signal rates are 
possible even after employing a strong photon angular cut. For example, if 
$m_1=170$ GeV with $|\theta_\gamma|>15^o$, then $E_\gamma=27.75$ GeV and 
$\sigma=0.3$ pb at $\sqrt s$=200 GeV and thus a 200 $pb^{-1}$ sample would 
yield 60 events which should be observable above the continuum background.

\nn

\subsubsection{Resonance Production at Future Colliders}

It is more likely that the first graviton KK state will be 
several hundreds of GeV or more in mass and we now explore the phenomenology
of this scenario in more detail than given in our previous work\cite{dhr}.
The basic signature for the RS model with the SM fields being confined to
the TeV-brane is the direct resonance production of the graviton KK 
excitations.  If it is kinematically feasible to produce more than one
KK tower member, the fact that the excitation spacing is proportional to
the root of the $J_1$ Bessel function provides a smoking gun signal for
the non-factorizable geometry of this model.  In addition, the two model
parameters which govern the 4-d phenomenology, \ie, $k$ and \lpi, can be 
completely determined\cite{dhr} by the measurement of the mass and width
of the first excitation.

\begin{figure}[htbp]
\centerline{
\psfig{figure=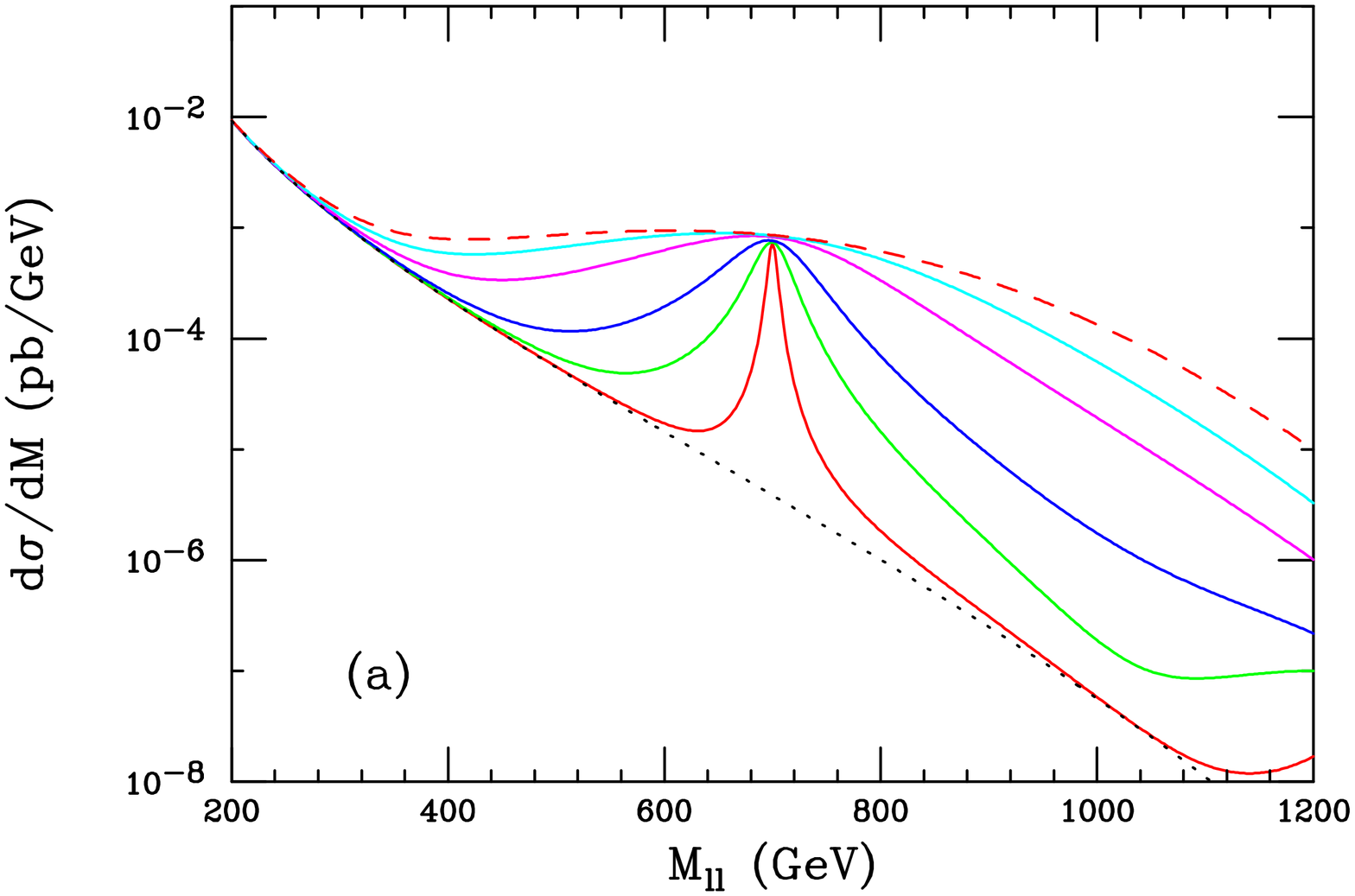,height=9.cm,width=12cm,angle=90}}
\vspace{0.1cm}
\centerline{
\psfig{figure=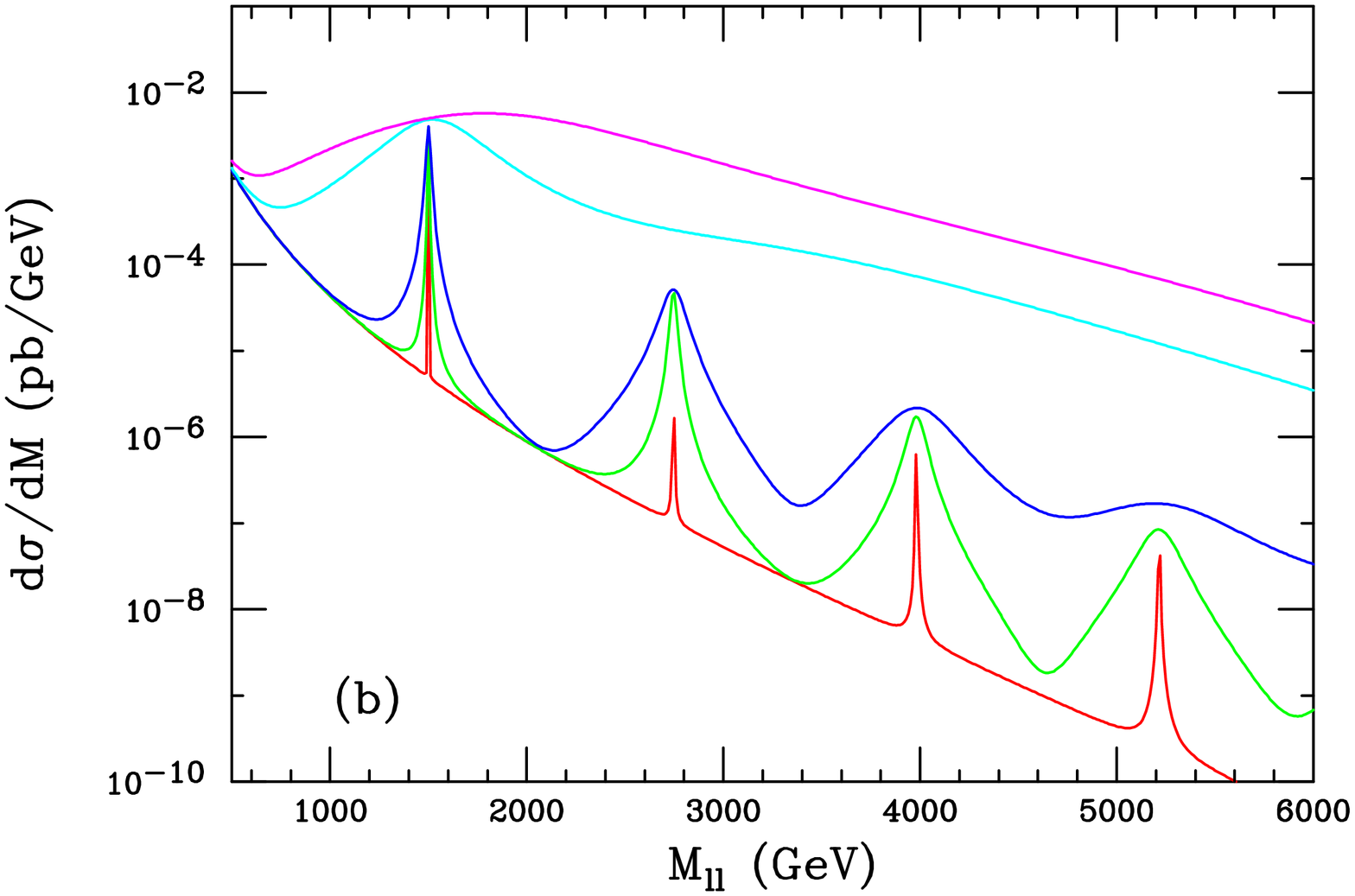,height=9.cm,width=12cm,angle=90}}
\caption{Drell-Yan production of a (a) 700 GeV KK graviton at the Tevatron
with $\cc=1,\, 0.7,\, 0.5,\, 0.3,\, 0.2$, and 0.1, respectively, from
top to bottom; (b) 1500 GeV KK graviton and its 
subsequent tower states at the LHC.  From top to 
bottom, the curves are for $\cc=1,\, 0.5,\, 0.1,\, 0.05$, and 0.01, 
respectively.}
\label{fig17}
\end{figure}

We first examine the cleanest signal for graviton resonance production,
namely an excess in Drell-Yan events from 
$q\bar q,gg\to G^{(n)}\to \ell^+\ell^-$.  The Drell-Yan line-shape is
presented in Fig. \ref{fig17} as a function of the invariant mass of
the lepton pair for $m_1^{\rm grav}=700,\, 1500$ GeV at
the Tevatron and LHC, respectively, for various values of \cc.  The production
of subsequent tower members are also shown for the LHC, note the increasing
widths of the higher resonances.  Also note that the value of the peak
cross section for the first resonance is independent of the value of \cc.  
We see that for
larger values of \cc, \eg, $\cc\gsim 0.5$, the bump structure of the
resonances is lost due to the large value of its width (recall that the
width is proportional to $[\cc]^2$) and the interference from the higher
excitations.  In this case, graviton production appears as a shoulder on
the SM predicted Drell-Yan spectrum, and is similar to the effect of contact 
interactions.  Nonetheless, we find that the resulting search reach for the 
first graviton excitation from a full calculation is essentially equivalent
to our earlier results\cite{dhr} where we employed 
the narrow width approximation.  These results are given
as a function of \cc\ in our previous work and are not reproduced here
with the exception that the results for run II
at the Tevatron with 2 \infb\ of integrated luminosity are displayed
in Fig. \ref{fig13}.

\begin{figure}[htbp]
\centerline{
\psfig{figure=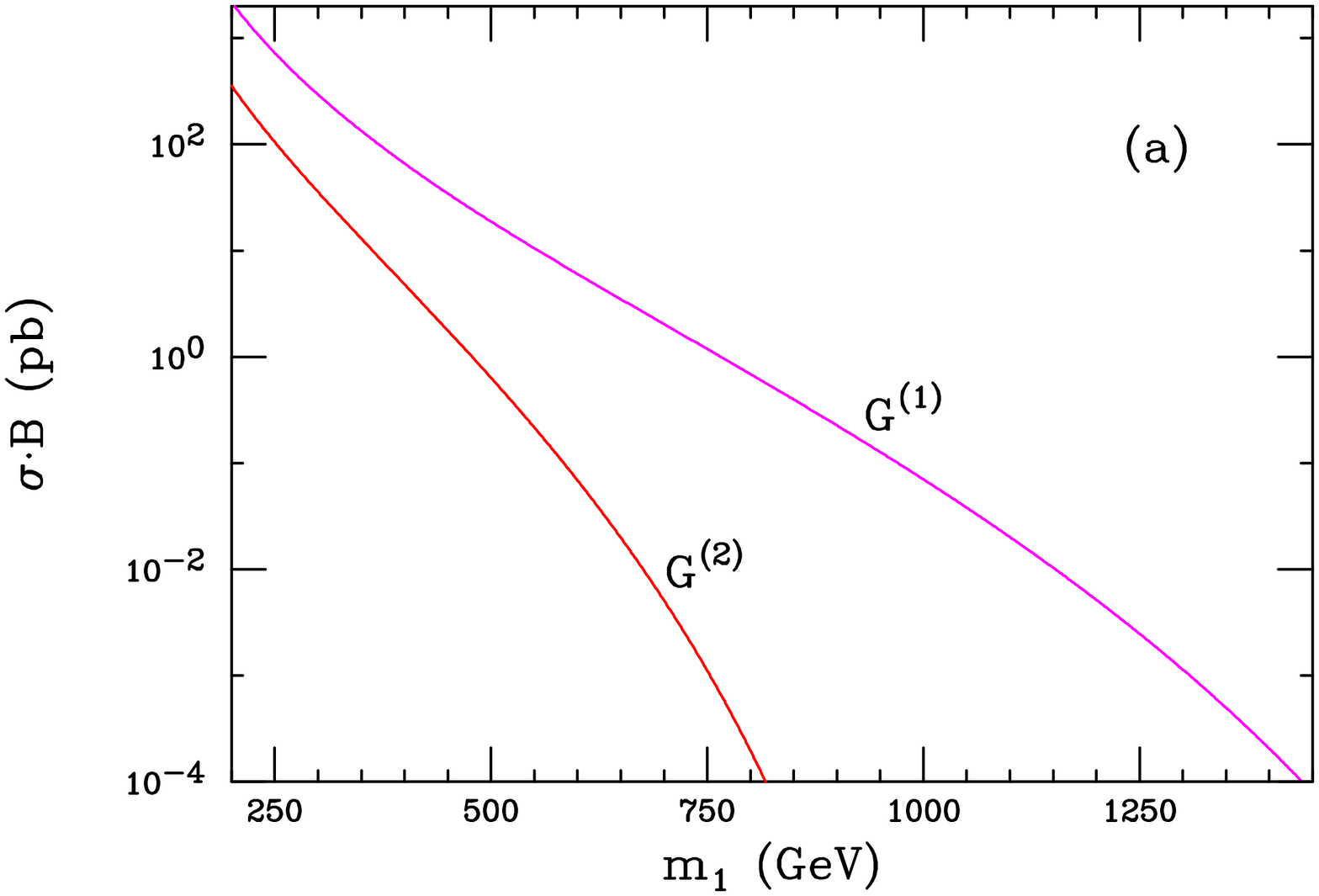,height=9.cm,width=12cm,angle=90}}
\vspace{0.1cm}
\centerline{
\psfig{figure=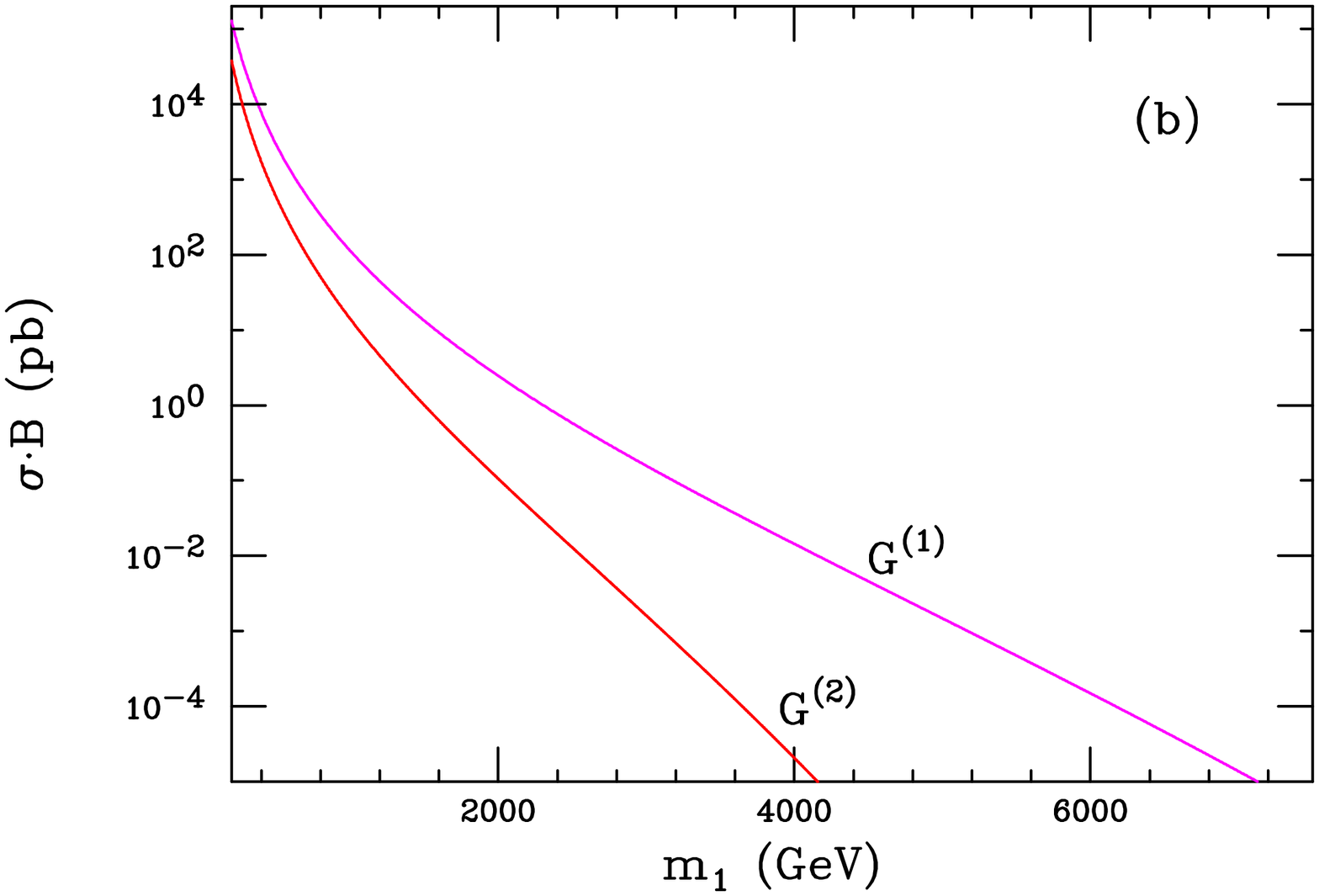,height=9.cm,width=12cm,angle=90}}
\caption{Cross sections for Drell-Yan production at the (a) Tevatron and 
(b) LHC of the first 
two graviton KK states coupling to the SM on the wall as a function of 
$m_1$. The upper (lower) curve in each case is for the first (second) KK state.
Here, we have set $\cc=0.1$.}
\label{fig18}
\end{figure}

Since the fundamental signature of a non-factorizable geometry
is the non-uniform spacing
of the graviton KK states, it is important to examine the probability of
observing the second excitation if the first resonance is discovered.  In
order to quantify this we show in Fig. \ref{fig18} the cross section times 
leptonic branching fraction
for the Drell-Yan production of the first two graviton KK states as a
function of the first excitation mass for the sample value $\cc=0.1$.  
We see that the second excitation has 
a sizable cross section at both accelerators.  We estimate that the $n=2$
graviton KK state will be discovered at the Tevatron (LHC) with 2 \infb\
(100 \infb) of integrated luminosity if the mass of the first excitation is
less than 725 GeV (3.8 TeV).  This is clearly a significant discovery reach.

\begin{figure}[htbp]
\centerline{
\psfig{figure=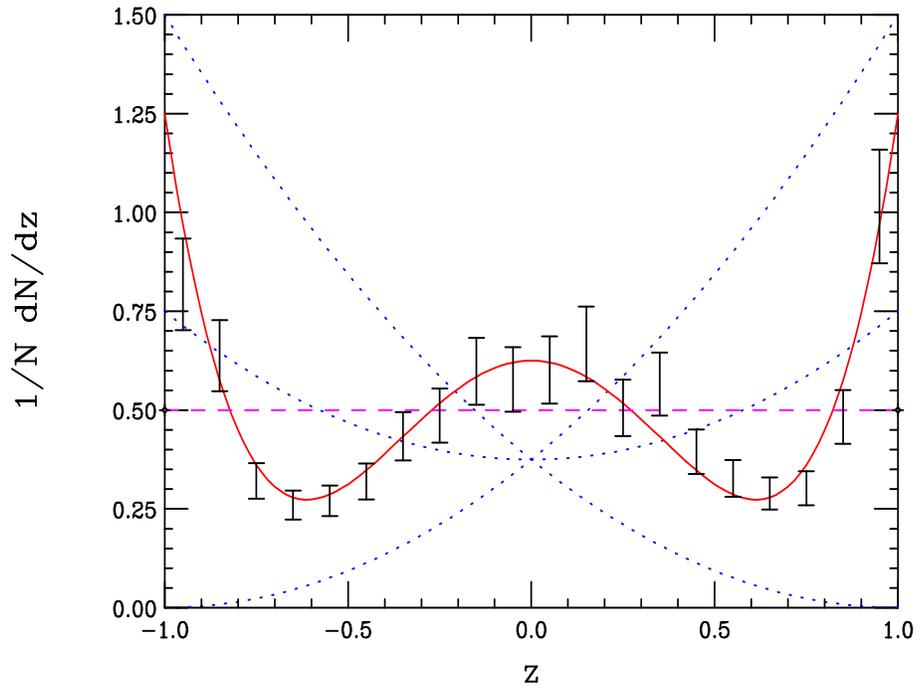,height=9cm,width=12cm,angle=90}}
\caption{Normalized angular distribution ($z=\cos\theta$) 
for the decay of a spin-2 graviton 
into fermion pairs (the `w'-shaped curve) in comparison to similar decays by 
either spin-0 (dashed) or spin-1 (dotted) particles. The data with errors show 
the result from a typical sample of 1000 events.}
\label{fig19}
\end{figure}

Next, we examine the ability of a hadron collider to determine the spin of
a new resonance once one is discovered.  It is well-known that the angular
distribution of a particle's decay products convey information about
its spin quantum number.  This is depicted in Fig. \ref{fig19} for the
decay of particles of various spins into fermion pairs.  We see that a
spin-0 resonance has a flat angular distribution, of course, spin-1 corresponds
to a parabolic shape, and spin-2 yields a quartic distribution.  The ability
of a collider to distinguish between these distributions depends on the
amount of available statistics.  For purposes of demonstration, 
we have generated the angular distribution, including statistical errors, of 
a typical data sample of 1000 events; this is displayed in Fig. \ref{fig19}.  
We see that with this level of statistics, the
spin-2 nature of a KK graviton is easily determined.  From Fig. \ref{fig18},
we see that the accumulation of 1000 events or more corresponds to a
value of $m_1^{\rm grav}\lsim 4200$ TeV with $\cc=0.1$ at the LHC with 
100 \infb\ of integrated luminosity.  Further study, similar to what has
been performed in the case of a new $Z$ boson resonance\cite{tgr},
is required in order to 
determine the range of parameter space for which the spin-2 nature of the
graviton can be resolved.

\begin{figure}[htbp]
\centerline{
\psfig{figure=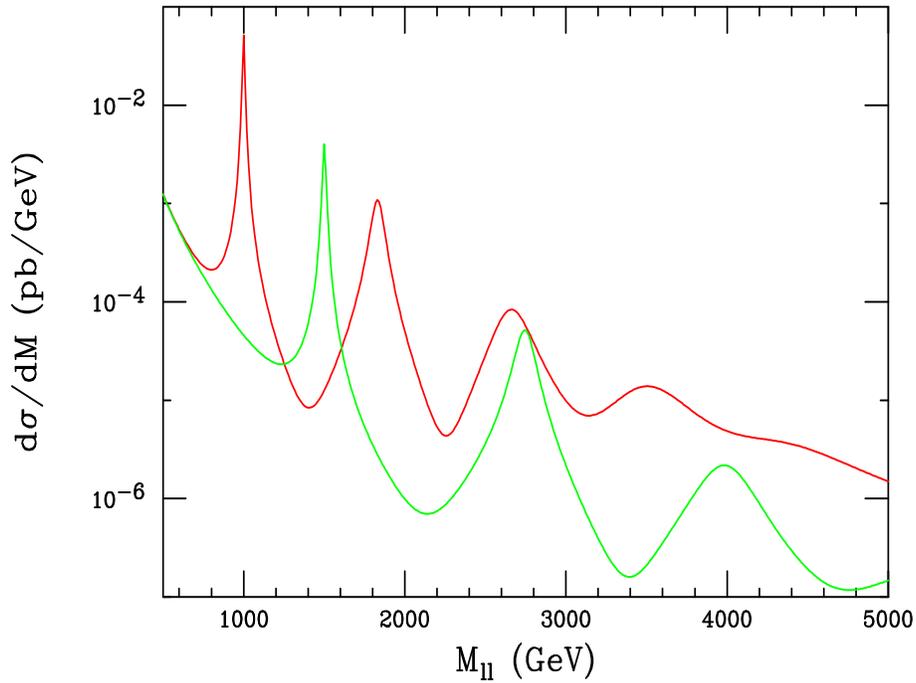,height=9cm,width=12cm,angle=90}}
\caption{Multiple KK graviton resonances produced at the LHC with 
$m_1^{\rm grav}=1$ TeV 
and $\cc=0.1$ and for $m_1^{\rm grav}=1.5$ TeV with $\cc=0.2$.}
\label{fig20}
\end{figure}

\begin{figure}[htbp]
\centerline{
\psfig{figure=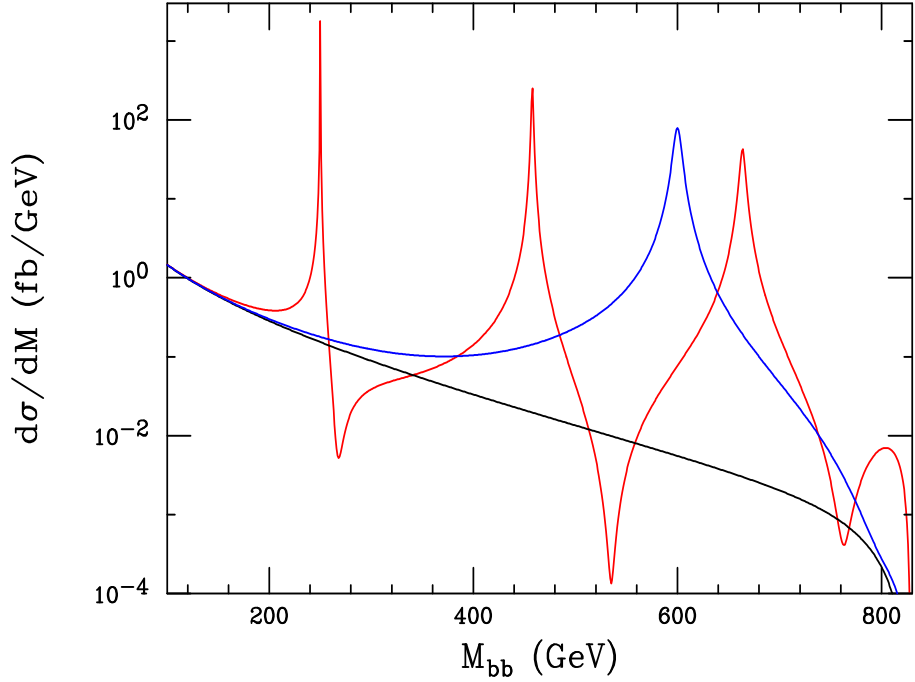,height=9cm,width=12cm,angle=90}}
\caption{$\gamma\gamma \to b\bar b$ showing graviton resonances assuming 
$m_1^{\rm grav}=250$ GeV and $\cc=0.03$ or with 
$m_1^{\rm grav}=600$ GeV and $\cc=0.1$.  The flat curve
corresponds to the expected SM background.}
\label{fig21}
\end{figure}

Lastly, we present the graviton KK spectrum with varied values of the
parameters in two sample processes.  The invariant mass spectrum of the
lepton pair is shown in Fig \ref{fig20} for Drell-Yan production of the
graviton KK spectrum at the LHC, comparing $m_1^{\rm grav}=1$ TeV with
$\cc=0.1$ with $m_1^{\rm grav}=1.5$ TeV with $\cc=0.2$.  Figure \ref{fig21}
displays the KK line-shape in $\gamma\gamma\to b\bar b$, comparing
$m_1^{\rm grav}=600$ GeV with $\cc=0.1$, $m_1^{\rm grav}=250$ GeV with
$\cc=0.03$, and the SM prediction.  These figures demonstrate how the
KK spectrum changes in terms of size of the peak cross sections and widths
of the resonances as the model parameters are varied.  These processes
were chosen simply for demonstration and for ease of identifying the final
state.  We emphasize that graviton KK resonance production will occur at
all planned colliders, and that the gravitons will decay into all possible
2-body final states with the relative branching fractions as given in
Fig. \ref{fig14}.  Observation of the relative rates of all these processes
would serve as an additional verification of the model.

\nn

\section{Conclusions}

In this paper we have explored the detailed phenomenology of the 
Randall-Sundrum model of localized gravity for the cases where the
SM field content propagates in the bulk or lies on the TeV-brane.  We have
derived the wavefunctions and interactions of the KK tower for each field
that is allowed to exist in the bulk.
We presented an argument demonstrating that if
spontaneous symmetry breaking takes place in the bulk, either the couplings
of the gauge bosons do not take their SM values, or the SM mass relationship
between the $W$ and $Z$ becomes corrupted, depending on whether the matter
fields exist in the bulk or not.
We thus conclude that the Higgs field must be confined to the TeV-brane.

In the scenario where the SM gauge and matter fields propagate in the extra 
dimension, our results can be summarized as:

\begin{itemize}

\item The phenomenology in this case is now governed by three parameters,
$k$, \lpi, and the bulk mass parameter, $\nu$.

\item  We found that the couplings of the resulting KK states are highly
dependent on the value of the bulk mass parameter.  We then identified four
regions with distinct phenomenologies, corresponding to different ranges of
$\nu$.

\item  We examined the phenomenological signatures of this model in all
four regions.  We compared the constraints placed on the model from precision
electroweak data with those obtainable from direct collider searches.
We found that the KK states couple too weakly in order to yield observable
signatures for $\nu< -0.5$.  The precision electroweak constraints resulted
in strong bounds for larger values of $\nu$
and indicate that the gauge and graviton KK states will not be kinematically 
accessible at the LHC for $\nu\gsim -0.3$.  In this case, the presence of
the KK towers will be probed via contact interaction searches.

\item  We also presented theoretical arguments for limiting the range of $\nu$.
We reasoned that $\nu\gsim -0.8$ to $-0.9$ in order to ensure that the 
fermion Yukawa couplings are not overly fine-tuned.  In addition, we saw that
$\nu$ cannot grow too large or else the precision electroweak bounds
translate into a value of \lpi\ which is far above the weak scale,
rendering the RS model irrelevant to the hierarchy problem.

\item Combining these theoretical and experimental constraints yields
a narrow range of $\nu$, $-0.9$ to $-0.8\leq\nu\leq -0.3$,  
for which the RS model
is viable and can be probed directly in colliders.  

\end{itemize}

This argues for a model that either selects $\nu$ to be in this narrow
viable range or prefers that the SM field content be constrained to lie
on the TeV-brane.

We thus also investigated the phenomenology of the RS model in this second 
case, expanding on our previous work.  In this case, gravity is the only
field which propagates in the extra dimension and expands into a KK
tower upon compactification.  The phenomenology is now governed by only
two parameters, with the fermion bulk mass obviously being absent.
We examined the possibility of lighter gravitons, which may be produced 
at LEP II as a direct resonance or in an emission process.
We computed the effects of the graviton KK states on the precision 
electroweak oblique parameters and found constraints on
the parameter space which are complementary to those obtainable from
direct collider searches.  In addition, we delineated the signatures for
the graviton KK spectrum at future colliders.

\begin{figure}[htbp]
\centerline{
\psfig{figure=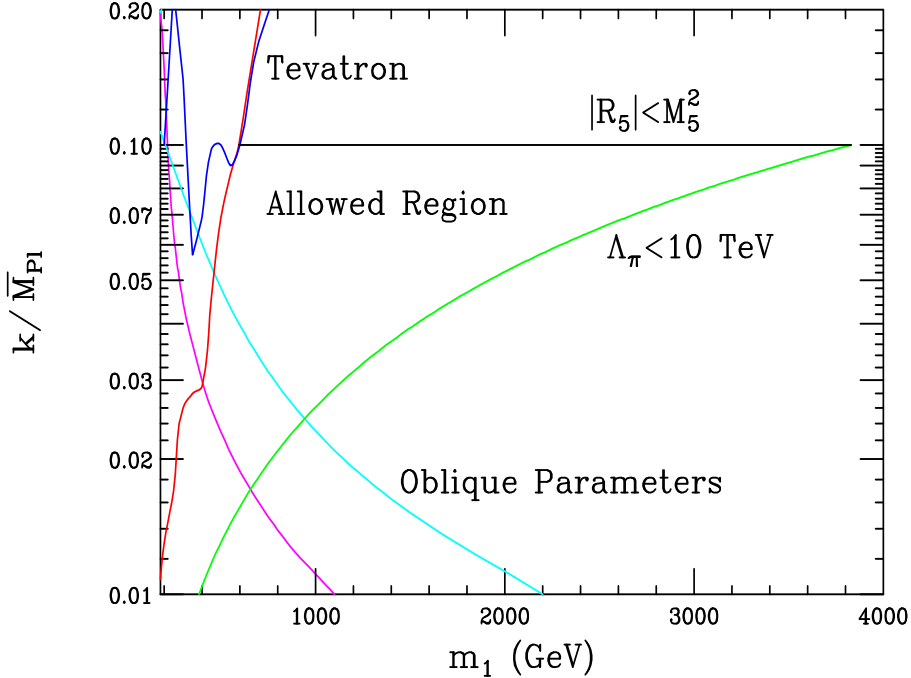,height=9cm,width=12cm,angle=90}}
\caption{Summary of experimental and theoretical constraints on the RS model,
for the case where the SM lies on the TeV-brane,
in the \cc\ and $m_1^{\rm grav}$ plane.  The allowed region lies in the
center as indicated.}
\label{fig22}
\end{figure}

The combined results of our analysis in the scenario where the SM fields
lie on the TeV-brane are presented in the parameter plane 
$\cc - m_1^{\rm grav}$ in  Fig. \ref{fig22}.  The constraints from present
data are summarized by the bounds from Drell-Yan and di-jet production at
the Tevatron from Run I and from the global fit to the 
oblique parameters $S$ and $T$, as labeled
in the figure.  In each case, the excluded area lies to the left of the
curves.  The theoretical constraints are given by curvature bound
$|R_5|=20k^2<M_5^2$, which yields $\cc<0.1$, and by the prejudice that
$\lpi\lsim 10$ TeV to ensure that the model resolves the hierarchy.  We see
that this synthesis of experimental and theoretical constraints results
in a small, {\it closed} allowed region in the model parameter space.
Comparing this allowed region with our previous results\cite{dhr} for the 
search reach for graviton production via the Drell-Yan mechanism at the LHC,
we see that the LHC will be able to cover this entire region of parameter
space with 100 \infb\ of integrated luminosity.  Hence, in the scenario
where the SM fields lie on the TeV-brane, the LHC will be able to 
definitively discover or exclude the RS model of localized gravity, if it
is relevant to the hierarchy.

\noindent{\bf Acknowledgements}:

We would like to thank N. Arkani-Hamed, Y. Grossman, M. Schmaltz, and 
R. Sundrum for discussions related to this work.  Thank you for reading this
paper and have a nice day.

\vspace{1in}

\noindent\appendix{\Large \bf Appendix A}

In this Appendix we will supply a robust argument against spontaneous 
symmetry breaking (SSB) by Higgs bosons in the RS Bulk. We assume that SSB 
takes place either in the bulk or on the wall so that if SSB in the bulk is 
untenable we are forced to consider the 
Higgs to lie only on the SM brane. Since 
there are no massless gauge 
KK modes when there are bulk gauge masses, we would 
now be forced to identify the SM $W$ and $Z$ bosons as the lowest massive 
KK modes of their respective towers. On the other hand the photon and gluons,
having no corresponding bulk mass terms, can be identified 
with the ordinary massless modes. 

To proceed we first consider the SM-like part of the action involving only 
the gauge and Higgs fields taking $y=r_c\phi$:
\be
S_{SM}=\int d^4xdy\sqrt{G}\left[-\sum_a {1\over 4}F^a_{MN}F^{MN}_a
+|D_A\phi |^2-V(\phi )+...\right]\,,
\ee
and follow all of the usual steps of SSB associated with the SM. The only 
difference with the usual result will be the labelling on the 5-d couplings 
and the Higgs vacuum expectation value (vev),
\ie, $g,g',e\to g_5,g_5',e_5$ and $v\to v_5$ etc. In the 
usual basis this generates bulk mass terms associated with the $Z$ and $W$ 
fields, 
\mbox{\boldmath $M_{Z,W}$} but none for the photon and gluon fields due to the 
remaining unbroken gauge invariance. We expect that both of these generated 
masses are naturally of order $k$ and that they are also related, assuming 
spontaneous symmetry breaking 
via Higgs doublets in the bulk, by the usual SM-like relationship  
$\mbox{\boldmath $M_{W}^2$}=\mbox{\boldmath $M_{Z}^2$}\cos^2 
\theta_5$ with, as usual,  
$g'_5/g_5=\tan \theta_5$, $\theta_5$ being the angle diagonalizing the 
$Z-\gamma$ mixing matrix. The 5-d coupling of the photon  is then 
identified as $e_5=g_5 \sin \theta_5$. Now although this all seems trivial and 
straightforward problems begin to appear when we try to match these 5-d 
couplings and the generated  masses to those in the usual 4-d SM.

Let us first consider the case where the SM fermions are in the bulk. Then, 
since the photon has no bulk mass term, it is easy to calculate the 
relationship between $e_5=g_5 \sin \theta_5=g_5s_5$ and $e=g \sin \theta=gs$
by considering the coupling between fermionic zero-modes, which we identify 
as the SM fields, with the photon tower zero-mode, \ie, the ordinary photon 
which has a constant wave function in the extra dimension. We obtain the
familiar relation
\be
e={e_5\over\sqrt{2\pi r_c}}\,\quad\quad\quad{\mbox or}\quad\quad\quad
{g_5s_5\over\sqrt{2\pi r_c}}=gs\,.
\ee
As discussed above, the $Z$ and $W$ of the SM are now identified with the 
lightest massive modes of their respective towers with wave functions of the 
form 
\be
\chi_{W,Z}={e^\sigma\over N_{W,Z}}\left[ J_{\alpha_{W,Z}}+\beta_{W,Z}
Y_{\alpha_{W,Z}}\right]\,,
\ee
where $N_{W,Z}$ is a normalization factor, $\beta_{W,Z}$ are constants, and
\be
\alpha_{W,Z}=\left[1+\mbox{\boldmath $M_{W,Z}$}^2/k^2\right]^{1/2}\,,
\ee
respectively.
Denoting the complete
fermion zero-mode wave functions symbolically by $f_\nu$ the 
relationship between the 5-d $W$ coupling and that for the SM is given by 
\be
{g_5\over\sqrt 2}\int dy \sqrt{G} f_\nu^2\chi_W\equiv {g_5\over\sqrt 2}I_W=
{g\over\sqrt 2}\,,
\ee
where $I_W$ represents the $y$ integration over the various wave functions. 
Note that we have assumed that {\it all} fermion flavors have the same value 
of $\nu$. If this were not the case universality violation would be rampant. 
In the $Z$ case, due to the structure of the coupling, we arrive at two 
necessary conditions for the correct matching
\bea
(i) & & {g_5\over c_5}\int dy \sqrt{G} f_\nu^2\chi_Z\equiv {g_5\over c_5}I_Z
={g\over c}\,,\\
(ii) & & {g_5\over c_5}s_5^2\int dy \sqrt{G} f_\nu^2\chi_Z\equiv {g_5\over c_5}
s_5^2I_Z={g\over c}s^2\,,\nonumber
\eea
where $I_Z$ represents the corresponding $y$ integration over the $Z$
and fermion wave functions. 
Dividing Eq. (50ii) by (50i), we arrive at $s_5=s$. Substituting Eq. (49) 
into Eq. (46) and using this $s_5=s$ result we arrive at the requirement that 
$I_W=1/{\sqrt {2\pi r_c}}$, independent of $\nu$ or 
$\mbox{\boldmath $M_W$}/k$!  This is of course in general 
impossible so we must conclude that if fermions are in the bulk the SSB 
breaking by bulk Higgs fields does not allow us to simultaneously 
recover the correct SM 
couplings for the photon, $W$ or $Z$. 

Now if the fermions are on the wall it is easy to see that $s_5=s$ and 
$g=g_5/\sqrt{2\pi r_c}$ are automatically 
consistent with all of the required coupling 
relations since we must evaluate the $W$ and $Z$ wave functions on the SM 
brane via delta functions. 
However now a different problem arises with the $W$ and $Z$ masses 
since we now require $x_{1W}=x_{1Z}\cos \theta$ where the $x_1$'s are the 
lowest roots of the appropriate combination of 
boundary condition equations that 
yield the tower mass eigenvalues. 
Furthermore we require that this condition must hold without any fine-tuning 
of the ratio $\mbox{\boldmath ${M_Z}$}/k$. 
To show that this condition does not hold naturally,
let us take as an example $\mbox{\boldmath $M_Z$/k}=1(2)$ from which we can 
calculate $x_{1W}^2/x_{1Z}^2=\cos ^2 \theta$; we find that $\cos^2\theta=
0.9359(0.8781)$ assuming that $\mbox{\boldmath $M_W$}=
\mbox{\boldmath $M_Z$}\cos \theta$ 
with $\cos \theta=0.77$ as input. 
Knowing the input values of both $\mbox{\boldmath $M_{Z}^2$}/k^2$ and
$\cos \theta$, which takes a common value in the bulk and on the wall, we can
fix the ratio $\mbox{\boldmath $M_{W}^2$}/k^2$. This then allows us to
evaluate the quantities $\alpha_{W,Z}$, as given by Eq.(48), which are the
indices of the Bessel functions for the $Z$ and $W$ tower member wave
functions in Eq.(47). Applying the usual $Z_2$-even boundary conditions on
these wave functions as discussed above we can determine the mass eigenvalues
for the lightest members of each of these towers that we are now identifying
with the $W$ and $Z$. The ratio of these eigenvalues should return 
the input value of $\cos \theta$ to us since $x_{1W}/x_{1Z}=\cos \theta$. If
we do not obtain the input value or we find that that the result depends on
the input value of $\mbox{\boldmath $M_{Z}$}/k$ we can conclude that this
approach is internally inconsistent. 
Since our input and output values are significantly different, we can conclude
that this possibility fails as well.
Thus if fermions are on the wall we may recover the 
correct SM couplings but the SM mass relationship between the $W$ and $Z$ 
becomes corrupted. This implies that the Higgs cannot generate SSB in the bulk 
when the fermions are on the SM brane.  Combining both arguments, we 
thus conclude from this discussion 
that SSB must take place on the SM brane and that therefore the Higgs fields 
are to be found there as well.

\vspace{1in}

\noindent\appendix{\Large \bf Appendix B}

In this Appendix we present concise expressions for the most common couplings 
discussed in the main text in the scenario where the fermion fields reside in 
the bulk.  The ${n^{\rm th}}$ 
graviton and gauge boson KK couplings to a pair of 
zero-mode SM fields are given in terms of simple 
integrals by:

\noindent{\mbox{\boldmath $f^{(0)}\bar f^{(0)}A^{(n)}$}:}
\be
C^{f\bar fA}_{00n}={g^{(n)}\over g^{SM}}=\sqrt{2\pi kr_c}\left[{1+2\nu\over 1-
\epsilon^{2\nu+1}}\right]\int^1_\epsilon dz ~z^{2\nu+1}
{J_1(x^A_nz)+\alpha^A_n Y_1(x^A_nz)\over |J_1(x^A_n)+\alpha^A_n Y_1(x^A_n)|}\,,
\ee

\noindent{\mbox{\boldmath $f^{(0)}\bar f^{(0)}G^{(n)}$}:}
\be
C^{f\bar fG}_{00n}={1\over\epsilon}\left[{1+2\nu\over 1-
\epsilon^{2\nu+1}}\right]\int^1_\epsilon dz ~z^{2\nu+2}{J_2(x^G_nz)\over
|J_2(x^G_n)|}\,,
\ee

\noindent{\mbox{\boldmath $A^{(0)}A^{(0)}G^{(n)}$}:}
\be
C^{AAG}_{00n}={1\over\epsilon} ~~{2(1-J_0(x^G_n))\over \pi kr_c(x_n^G)^2
|J_2(x^G_n)|}\,,
\ee
where $\alpha^A_n$ is defined in Eq. (14), $\epsilon\equiv e^{-kr_c\pi}$, and
the $x_n^{A,G}$ denote the appropriate Bessel roots that appear in the gauge 
and graviton KK wavefunctions as given in Section 2.
Note that the coupling of two zero-mode gauge bosons to the $n^{\rm th}$ KK 
graviton can be computed analytically.  In a similar manner we find the
following expressions for couplings involving only a single zero-mode SM
field:

\noindent{\mbox{\boldmath $f^{(\ell)}\bar f^{(0)}A^{(n)}$}:}
\be
C^{f\bar fA}_{\ell 0n}=
\sqrt{2\pi kr_c}\left| {2(1+2\nu)\over 1-\epsilon^{2\nu+1}}
\right|^{1/2}\int^1_\epsilon dz ~z^{\nu+3/2}\, {J_f(x^L_\ell z)\over |J_f(
x^L_\ell)|}\, {J_1(x^A_n z)+\alpha_nY_1(x^A_nz)\over |J_1(x^A_n)
+\alpha_nY_1(x^A_n)|}\,,
\ee

\noindent{\mbox{\boldmath $f^{(\ell)}\bar f^{(0)}G^{(n)}$}:}
\be
C^{f\bar fG}_{\ell 0n}
={1\over\epsilon}\left| {2(1+2\nu)\over 1-\epsilon^{2\nu+1}}
\right|^{1/2}\int^1_\epsilon dz ~z^{\nu+5/2}\, {J_f(x^L_\ell z)\over 
J_f(x^L_\ell)}\,
{J_2(x^G_n z)\over |J_2(x^G_n)|}\,,
\ee

\noindent{\mbox{\boldmath $A^{(\ell)}A^{(0)}G^{(n)}$}:}
\be
C^{AAG}_{\ell 0n}={2\over\epsilon\sqrt{2\pi kr_c}}\int^1_\epsilon 
dz ~z^2\, {J_1(x^A_\ell z)+\alpha_\ell^A Y_1(x^A_\ell z)\over 
|J_1(x^A_\ell)+\alpha_\ell^A Y_1(x^A_\ell)|}\,
{J_2(x^G_n z)\over |J_2(x^G_n)|}\,,
\ee
where $f=\nu-1/2\, (-\nu+1/2)$ for $\nu>\, (<) -1/2$, and $x_\ell^L$ correspond
to the Bessel roots for the Left-handed fermion KK tower.

A 4-point coupling, between $\ell^{th}$ fermion - $0^{th}$ fermion -
$0^{th}$ gauge - $n^{th}$ graviton, is also present and is given by:

\noindent{\mbox{\boldmath $f^{(\ell)}\bar f^{(0)}A^{(0)}G^{(n)}$}:}
\be
C^{f\bar fAG}_{\ell 00n}= {1\over\epsilon}
\left| {2(1+2\nu)\over 1-\epsilon^{2\nu+1}}
\right|^{1/2}\int^1_\epsilon dz ~z^{\nu+5/2}\, {J_f(x^L_\ell z)\over
|J_f(x^L_\ell)|}\, {J_2(x^G_nz)\over |J_2(x^G_n)|}\,,
\ee
which is exactly the same as  $C^{f\bar fG}_{\ell 0n}$.

Let us now turn to the wall Higgs couplings to zero-mode bulk fields starting 
from the action
\be
S_{ffH}={\tilde \lambda_5\over {k}} \int d^4xdy\sqrt{G}\bar\Psi(x,y) 
\Psi(x,y) H^0(x)\delta(y-r_c\pi)\,, 
\ee
where a factor of 
$k$ has been introduced to render $\tilde \lambda_5$ dimensionless. 
When the Higgs gets a vev of order the Planck scale, $v_5$, we must shift the 
field as $H^0\to v_5+H'^0$. If 
we substitute the fermion mode expansions and extract out the zero-mode 
pieces and let $H'^0\to \epsilon^{-1} H'$ 
to account for the required rescaling 
of the Higgs field kinetic term, we can identify the 4-d coupling as 
$\lambda_4=\tilde \lambda_5 \omega/2$ (with $\epsilon v_5=v_4$) using the
familiar ratio 
\begin{equation}
\omega= {(1+2\nu)\over {1-\epsilon^{1+2\nu}}}\,,
\end{equation}
which multiplies $v_4$ and which has important implications as discussed in 
the text. Note that $v_4$ is now naturally of order the TeV scale. 
One also finds that the off-diagonal mode Yukawa couplings are induced from the 
same action. For example, the coupling of the $n^{\rm th}$ and $m^{\rm th}$ 
non-zero 
tower members to the Higgs is found to be $\tilde \lambda_5 (-1)^{m+n}$ 
while the coupling of a zero-mode and an $n^{\rm th}$ mode fermion to the Higgs 
is given by $ \tilde \lambda_5 (-1)^n\sqrt {\omega/2} $.
Thus the fermion tower members are seen to mix with themselves with a strength 
that is characterized by the induced zero-mode mode mass, \ie, the mass of 
the corresponding SM fermion. For all SM fermions, except perhaps for the top 
quark, these effects are quite small since we expect that the unmixed tower 
fermion masses begin in the range of hundreds of GeV if not larger. A similar 
analysis of the $W$ and $Z$ tower shows that the wall Higgs field induces 
the correct photon, $W$ and 
$Z$ SM masses. Here we need to identify the 4-d and 
5-d gauge couplings through the usual 
relation $g_4=g_5/\sqrt {2\pi r_c}$ and as 
before make use of the rescaling $v_4=\epsilon v_5$. Again 
one finds that mixing between 
the gauge fields within these individual towers with a strength characterized 
by the induced mass of the zero-mode as occurs in non-warped 
space{\cite {sminbulk,getV}}. 

%
\def\IJMP #1 #2 #3 {Int. J. Mod. Phys. A {\bf#1},\ #2 (#3)}
\def\MPL #1 #2 #3 {Mod. Phys. Lett. A {\bf#1},\ #2 (#3)}
\def\NPB #1 #2 #3 {Nucl. Phys. {\bf#1},\ #2 (#3)}
\def\PLBold #1 #2 #3 {Phys. Lett. {\bf#1},\ #2 (#3)}
\def\PLB #1 #2 #3 {Phys. Lett.  {\bf#1},\ #2 (#3)}
\def\PR #1 #2 #3 {Phys. Rep. {\bf#1},\ #2 (#3)}
\def\PRD #1 #2 #3 {Phys. Rev.  {\bf#1},\ #2 (#3)}
\def\PRL #1 #2 #3 {Phys. Rev. Lett. {\bf#1},\ #2 (#3)}
\def\PTT #1 #2 #3 {Prog. Theor. Phys. {\bf#1},\ #2 (#3)}
\def\RMP #1 #2 #3 {Rev. Mod. Phys. {\bf#1},\ #2 (#3)}
\def\ZPC #1 #2 #3 {Z. Phys. C {\bf#1},\ #2 (#3)}
\def\EPJ #1 #2 #3 {Euro. Phys. Jour. {\bf#1},\ #2 (#3)}

\end{document}